\documentclass[traditabstract]{aa}

\usepackage{graphicx}
\usepackage{txfonts}
\usepackage{natbib}
\usepackage{longtable,pdflscape}
\usepackage{booktabs}
\usepackage{multirow}
\usepackage[breaklinks, colorlinks, citecolor=blue]{hyperref}
\usepackage[flushleft]{threeparttable}
\newcommand{\lsim}{{\;\raise0.3ex\hbox{$<$\kern-0.75em\raise-1.1ex\hbox{$\sim$}}\;}}
\newcommand{\gsim}{{\;\raise0.3ex\hbox{$>$\kern-0.75em\raise-1.1ex\hbox{$\sim$}}\;}}

\def\arcs{\ifmmode {^{\scriptstyle\prime\prime}}
          \else $^{\scriptstyle\prime\prime}$\fi}
\def\parcm{\sa=.08em \sb=.03em
     \ifmmode \hbox{\rlap{.}\kern\sa}^{\scriptstyle\prime}\hbox{\kern-\sb}
     \else \rlap{.}\kern\sa$^{\scriptstyle\prime}$\kern-\sb\fi}
\def\arcm{\ifmmode {^{\scriptstyle\prime}}
          \else $^{\scriptstyle\prime}$\fi}
\def\parcs{\sa=.07em \sb=.03em
     \ifmmode \hbox{\rlap{.}}^{\scriptstyle\prime\kern -\sb\prime}\hbox{\kern -\sa}
     \else \rlap{.}$^{\scriptstyle\prime\kern -\sb\prime}$\kern -\sa\fi}

\def\apjl{ApJL}

\begin{document}

\title{Unveiling the shape: A multi-wavelength analysis of the galaxy clusters Abell 76 and Abell 1307}
\titlerunning{}

\author{R.~Barrena \inst{1,2} \and L.~Pizzuti \inst{3} \and G. Chon \inst{4,5} \and H. B\"ohringer \inst{6,5,4}}
% \author{The cool-core research group}
\institute{Instituto de Astrof\'{\i}sica de Canarias, C/V\'{\i}a L\'{a}ctea s/n, E-38205 La Laguna, Tenerife, Spain\\
\email{rbarrena@iac.es} 
\and
Universidad de La Laguna, Departamento de Astrof\'{i}sica, E-38206 La Laguna, Tenerife, Spain
\and
Dipartimento di Fisica G. Occhialini, Universit\`a degli Studi di Milano Bicocca, Piazza della Scienza 3, I-20126 Milano, Italy
\and
Max-Planck-Institut f\"ur Physik, Boltzmannstr. 8, D-85748, Garching, Germany
\and
Universit\"ats-Sternwarte M\"unchen, Fakult\"at f\"ur Physik,
Ludwig-Maximilian-Universit\"at M\"unchen, Scheinerstr. 1, D-81679 M\"unchen,
Germany
\and
Max-Planck-Institut f\"ur extraterrestrische Physik, Gie\ss enbachstra\ss e 1, D-85748 Garching, Germany
}

\date{Received ; accepted } 

\authorrunning{Barrena et al.}

\abstract{We analyse the dynamical state of the galaxy clusters Abell 76 and Abell 1307 from the optical 
point of view, presenting a coherent scenario that responds to the X-ray emissions observed in these 
structures. Our study is based on 231 and 164 spectroscopic redshifts, for the clusters A76 and A1307, 
respectively, obtained mostly with the {\it Telescopio Nazionale Galileo}, and complemented with others 
collected from the SDSS DR16 spectroscopic database and the literature. We find that A76 and A1307 are
two galaxy clusters at $z=0.0390$ and 0.0815, respectively, with a velocity dispersion of $650 \pm 56$ km 
s$^{-1}$ and $863 \pm 85$ km s$^{-1}$, and they show velocity distributions following,  in practice, Gaussian 
profiles. From our dynamical analysis, X-ray studies and SZ-{\it Planck} emission, we obtain a mean total mass 
M$_{500} = 1.7 \pm 0.6 \cdot 10^{14}$ M$_{\odot}$ and $3.5 \pm 1.3 \cdot 10^{14}$ M$_{\odot}$ for A76 and 
A1307, respectively. Using the SDSS DR16 photometric database, we find that the spatial distribution of 
likely cluster members in the case of A76 is very anisotropic, while A1307 shows a compact distribution 
of galaxies, but it is double peaked and elongated in the south-north direction. Using {\it XMM-Newton} X-ray data, 
we compared the surface brightness maps with galaxy distributions and noticed that both distributions are 
correlated. We reconstructed the total mass profile and velocity anisotropy of both clusters by analysing 
the full projected phase space, through the {\tt MG-MAMPOSSt} code. Our study reveals a slight indication 
of radial orbits for A76, while A1307 seems to prefer more isotropic orbits in the whole cluster range. 
In summary, A76 represents a typical young cluster, in an early stage of formation, with a very low X-ray 
surface brightness but a high temperature showing a very anisotropic galaxy distribution. A1307 is however 
more consolidated and massive showing in-homogeneous galaxy distribution and an asymmetric X-ray emission, 
which suggest a scenario characterised by recent minor mergers.}

\maketitle

\section{Introduction}
\label{sec:intro}

The formation of galaxy clusters (GCs) is governed by the collapse of the largest gravitationally bound
overdensities in the initial density field. These massive structures are composed of dark matter, gas, and 
galaxies, which gives them an important role in the study of structure formation, galaxy evolution, 
thermodynamics of the intergalactic medium, and plasma physics. Furthermore, their large mass makes them 
an essential observational test of the growth of structure over cosmological time. So, the hierarchical 
models of cluster formation predict that mergers involving small galaxy systems play an important role in 
the cluster assembly and evolution, mainly in those systems showing low-medium mass (e.g. \citealt{Ber09}; 
\citealt{McG09}). Cluster merger processes are characterised by the fact that galaxies and gas react 
at different timescales. Hence the importance of studying galaxy clusters from a multi-wavelength point 
of view. While the galaxy component is mainly studied using optical photometric and spectroscopic 
observations, X-ray emission reflects the gas behaviour. In this paper, we analyse the dynamics of two 
nearby clusters, Abell 76 and Abell 1307 (hereafter A76 and A1307) in order to investigate their 
formation as low-medium mass clusters in the nearby Universe.

A76 is part of the Extended Northern ROSAT Galaxy Cluster Survey (NORAS II; \citealt{Boh17}) 
and named RXCJ0039.9+0650. It has been classified as a galaxy cluster showing one of the lowest X-ray 
surface brightnesses among the nearby GCs. Its X-ray emission has been extensively studied by \citet{Ota13} 
(hereafter OTA13) using {\it Suzaku} observations, which reveals a very clumpy shape in surface brightness 
and an extremly high entropy in the innermost region. Additionally, {\it XMM-Newton} observations were 
carried out by the {\it XMM-Newton} Cluster Survey DR1 (XCS-DR1; \citealt{Meh12}). These studies suggest 
that A76 is a young, low-mass GC, at an early stage of cluster formation. A76 was also detected by the 
{\it Planck} satellite as a Sunyaev-Zeldovich (SZ) source \citep{Sun69}, named PSZ2 G117.98-55.88, with 
an estimated SZ mass of M$_{\rm{SZ}}$=$1.7 \pm 0.2 \cdot 10^{14}$ M$_{\odot}$ \citep{PC27}. The 
Meta-Catalogue of X-ray detected Clusters of galaxies (MCXC) by \citet{Pif11} also compiles A76 as MCXC 
J0040.0+0649 and estimates a radius and mass of R$_{500}=0.70$ Mpc and M$_{500}=0.99 \cdot 10^{14}$ 
M$_{\odot}$, respectively. Finally, we mention that A76 was also detected in the Dark Energy Survey (DES), 
and in the Sloan Digital Sky Survey (SDSS) DR8 photometric data set, as galaxy overdensity 
(RMJ003956.0+065054.8) showing a M$_{500}>10^{14}$ M$_{\odot}$.

Similarly, A1307 is a GC named RXCJ1132.8+1428 in the NORAS/REFLEX sample \citep{Boh00}; MCXC 
J1132.8+1428 in the MCXC survey, showing a radius and mass R$_{500}=0.99$ Mpc and M$_{500}=3.0 
\cdot 10^{14}$ M$_{\odot}$, respectively; and PSZ2 G243.64+67.74 in the second {\it Planck} SZ source 
catalogue, with a M$_{SZ}$=$3.9 \pm 0.2 \cdot 10^{14}$ M$_{\odot}$. In addition, the optical follow-up of 
{\it Planck} SZ sources, performed by \citet{Agu22} estimates a dynamical mass of M$_{500}=2.6 \pm 0.5 \cdot 
10^{14}$ M$_{\odot}$. This system has not been explored in detail in X-rays, however, \citet{And16}, 
analysing the galaxy component, suggests certain indications of bimodality. 

A1307 is a CHEX-MATE\footnote{See \url{http://xmm-heritage.oas.inaf.it/}} target and 
has been recently observed by the {\it XMM-Newton} satellite. In this work, we present a detailed study
based on new spectroscopic redshift observations in order to disentangle the dynamics of these two low-medium
mass clusters and propose a coherent scenario in order to explain their X-ray emissions. With this aim in 
mind, we performed a detailed study of the mass profile and velocity anisotropy of the member galaxies 
in each cluster using the MG-MAMPOSSt code by \citet{Piz21}. This method allowed us to jointly reconstruct 
the velocity anisotropy profile, $\beta(r)$, and the cluster total mass profile, which is an excellent approach 
to obtain a realistic view of the galactic component behaviour of these systems, and compare it with 
the gas dynamics present in A76 and A1307.
 
This paper is organised as follows. In Sect. \ref{sec:optical_obs} we describe the new spectroscopic 
observations as well as the X-ray data. We analyse the optical and galaxy properties of A76 and A1307 
in Sect. \ref{sec:optical_ana}, and compare them with X-ray properties in Sect. \ref{sec:Xprop}. We 
present our estimation of the dynamical mass and a more detailed analysis on the mass profile and 
velocity anisotropy in Sects. \ref{sec:mass} and \ref{sec:mass_recon}, respectively, further discussing 
our findings in Sect. \ref{sec:optical_xray_mass}. We conclude this paper by summarising our results in 
Sect. \ref{sec:conclusions}. 

In this paper, we assume a flat cosmology with $\Omega_m=0.3$, $\Omega_\Lambda=0.7$, and H$_0=70$ 
$h_{70}$ km s$^{-1}$ kpc$^{-1}$. Under this cosmology, 1 arcmin corresponds to 46 and 92 $h_{70}^{-1}$ kpc 
at the redshift of A76 and A1307, at $z = 0.0390$ and 0.0815, respectively.

\section{Data sample}
\label{sec:optical_obs}

\subsection{Optical spectroscopy and photometry}
\label{sec:spectroscopy}

Even though A76 and A1307 have been observed in X-ray, radio frequencies and broad-band optical photometry, 
the spectroscopic information available in public databases and in the literature is not enough to
perform a detailed study of their galactic component and dynamics. We find 55 spectroscopic redshifts in 
the NASA/IPAC Extragalactic Database (NED) for A76, mainly obtained by \citet{Rin16} using the {\it Hectospec/MMT}
spectrograph as part of the HeCS-SZ survey. A1307 was spectroscopycally sampled in the SDSS-DR16 survey, 
and we find 45 cluster members within a region of $11^\prime$ radius with respect to the centre of the 
cluster. Therefore, in June 2020 and September 2023, in order to extend these samples, we performed 
spectroscopic observations at the 3.5m {\it Telescopio Nazionale Galileo} (TNG) telescope, at Roque de los 
Muchachos Observatory (La Palma), using the {\it DoLoRes} multi-object spectrograph.

The multi-object spectroscopic (MOS) observations allows us to obtain a large number of redshifts in a short
time and a limited field. Using this technique, we acquired spectroscopic redshifts covering a region of 
$27^\prime \times 22^\prime$ of A76 and a likely circular field of $11^\prime$ radius in A1307. The A76 field
was mapped using 12 MOS masks including more than 300 slitlets, obtaining good quality redshifts for 221
galaxy targets. Similarly, A1307 was covered with 5 masks containing about 140 slitlets and obtaining 121
spectra with good quality. The slitlets were placed in order to avoid the SDSS and NED redshift samples 
and so maximise the acquisition of new data. We used the DoLoRes spectrograph, installed in the 3.5m TNG 
telescope, in low resolution observing mode with the (low resolution) LR-B grism\footnote{See \url{http://www.tng.iac.es/instruments/lrs} 
for LR-B grism technical details}. Masks were manufactured with slits of $1.6\arcsec$ width, which provides 
a dispersion of 2.75 \AA $\ $ per pixel between 370 and 800 nm of wavelength coverage. We acquired a single 
frame per mask, with 1200 s exposure time in A76 and 1800 s in A1307, respectively. We processed the spectra 
using standard {\tt IRAF} packages and calibrated in wavelength using neon, mercury and helium arcs.

We estimated the spectroscopic redshift by correlating them with a set of templates retrieved from the Kennicutt 
Spectrophotometric Atlas of Galaxies (\citealt{ken92}). We applied the technique detailed by \citet{Ton79} 
and implemented as the task {\tt RVSAO.XCSAO} in {\tt IRAF} facility. This procedure detects and correlates 
the main features registered in the scientific spectra (i.e. the Ca H and K doublet, H$_\delta$, G band, and MgI 
in absorption, and the OII, OIII doublet, and H$_\alpha$ and H$_\beta$ lines in emission) with those present in 
the template ones. The set of template spectra includes five galaxy morphologies (elliptical, Sa, Sb, Sc, 
and Irr types) in order to match properly the different shapes of scientific data. This procedure provides
a radial velocity estimate and the corresponding correlation error for all spectra observed. We added
to our sample the spectroscopic redshift present in NED and SDSS-DR16 spectroscopic databases. Tables 
\ref{tab:catalog_A76} and \ref{tab:catalog_A1307} (the full version of these tables are available electronically 
at CDS; here we only present a few lines as example of their contain) list the full spectroscopic samples in the 
A76 and A1307 fields, which compile 231 and 164 spectroscopic redshifts, respectively. The new spectra obtained 
with the TNG present a mean signal-to-noise ratio (S/N) of 9.5 and a mean error of 57 km s$^{-1}$. We detected 
78 and 32 star-forming galaxies in A76 and A1307 fields, respectively. These galaxies are characterised by the 
existence of $[$OII$]$, $[$OIII$]$ and/or H$_\alpha$ emission lines, that we are able to detect if they have 
equivalent widths $>10$ \AA.

\begin{table}
\fontsize{9}{11}
% \begin{threeparttable}
\caption{Velocity catalogue of galaxies measured spectroscopically in the A76 field. }
\label{tab:catalog_A76}
\tiny
\begin{tabular}{l c c p{5mm} p{5mm} l}
\midrule \midrule
 ID & R.A. \& Dec. (J2000)              & v$\pm \Delta$v & \makebox[8mm][c]{$g^\prime$} & \makebox[8mm][c]{$r^\prime$} & Notes \cr
    & R.A.=$00\! : \! mm \! : \! ss.ss$ & (km s$^{-1}$)  &            &            & \cr
    & Dec.=$dd\! :\! mm \! : \! ss.s$  &                &            &            & \cr
\midrule
  001$^\star$ & 38:27.67 \enspace 6:39:33.4 & 10650  $\pm$ 24	& 16.21 & 14.94 & \cr      
  002$^\star$ & 38:28.86 \enspace 6:41:14.5 & 12329  $\pm$ 24	& 16.08 & 14.85 & \cr       
  003$^\star$ & 38:41.96 \enspace 7:17:21.1 & 12208  $\pm$ 24	& 15.83 & 14.58 & \cr      
  004$^\star$ & 38:45.82 \enspace 6:28:03.5 & 11529  $\pm$ 24	& 15.41 & 14.08 & \cr       
  005$^\star$ & 38:49.72 \enspace 6:46:42.6 & 11260  $\pm$ 24	& 15.68 & 14.39 & \cr        
  006$^\star$ & 38:52.78 \enspace 7:04:22.7 & 12489  $\pm$ 24	& 15.85 & 14.59 & \cr       
  007$^\star$ & 38:54.68 \enspace 7:03:45.7 & 12189  $\pm$ 24	& 14.30 & 12.93 & \cr       
  008$^\star$ & 38:54.73 \enspace 7:03:24.2 & 12100  $\pm$ 24	& 15.09 & 13.84 & \cr      
  009	        & 38:54.90 \enspace 7:02:53.2 & 8215   $\pm$ 24	& 15.78 & 14.52 & \cr        
  010$^\star$ & 39:01.69 \enspace 6:48:51.7 & 13024  $\pm$ 24	& 15.54 & 14.68 & \cr       
  011$^\star$ & 39:06.12 \enspace 6:52:22.0 & 12138  $\pm$ 57	& 17.30 & 16.22 & \cr  
  012         & 39:05.03 \enspace 6:42:45.8 & 14604  $\pm$ 24	& 16.39 & 15.19 & \cr          
  013	      & 39:06.32 \enspace 6:54:28.4 & 85220  $\pm$ 100  & 20.75 & 18.35 & ELG \cr     
... & ... \ \ \ \ \   \enspace \ \ \ \ \ \ ... &  \ \ ... &  \ \ \  ... & \ \ \ ... & ... \cr

\midrule
\end{tabular}
% \end{threeparttable}
\footnotesize{\textbf{Note 1:} asterisk in column 1 (ID) indicates the galaxies selected as cluster members.}  \newline
\footnotesize{\textbf{Note 2:} the full version of this table is available electronically at the CDS.}
\end{table}

\begin{table}
\fontsize{9}{11}
% \begin{threeparttable}
\caption{Velocity catalogue of galaxies measured spectroscopically in the A1307 field.}
\label{tab:catalog_A1307}
\tiny
\begin{tabular}{l c c p{5mm} p{5mm} l}
\midrule \midrule
 ID & R.A. \& Dec. (J2000)              & v$\pm \Delta$v & \makebox[8mm][c]{$g^\prime$} & \makebox[8mm][c]{$r^\prime$} & Notes \cr
    & R.A.=$11\! : \! mm \! : \! ss.ss$ & (km s$^{-1}$)  &            &            & \cr
    & Dec.=$+14\! :\! mm \! : \! ss.s$  &                &            &            & \cr
\midrule
 001$^\star$ & 32:08.49 \enspace 31:03.3 & 22583  $\pm$ 4   & 17.77 & 16.93 & \cr      % 524$^\star$
 002$^\star$ & 32:14.51 \enspace 25:07.4 & 24759  $\pm$ 5   &  --   &  --   & \cr      % 523$^\star$
 003	     & 32:14.90 \enspace 25:03.6 & 114880 $\pm$ 100 &  --   &  --   & ELG \cr  % 408	  
 004$^\star$ & 32:15.91 \enspace 30:40.7 & 24684  $\pm$ 4   & 17.32 & 16.40 & \cr      % 525$^\star$
 005$^\star$ & 32:18.67 \enspace 28:47.5 & 25550  $\pm$ 100 & 19.74 & 19.42 & ELG \cr  % 433$^\star$
 006$^\star$ & 32:20.61 \enspace 24:26.6 & 25685  $\pm$ 60  & 19.10 & 18.23 & \cr      % 404$^\star$
 007	     & 32:21.42 \enspace 29:26.5 & 6585   $\pm$ 149 & 19.63 & 19.02 & \cr      % 438	  
 008	     & 32:23.09 \enspace 28:55.1 & 62700  $\pm$ 100 & 20.54 & 19.84 & ELG \cr  % 434	  
... & ... \ \ \ \ \   \enspace \ \ \ \ \ \ ... &  \ \ ... &  \ \ \  ... & \ \ \ ... & ... \cr
\midrule
\end{tabular}
% \end{threeparttable}
\footnotesize{\textbf{Note 1:} asterisk in column 1 (ID) indicates the galaxies selected as cluster members.} \newline
\footnotesize{\textbf{Note 2:} The full version of this table is available electronically at the CDS.}
\end{table}
% \end{landscape}

In the A76 field, we observed repeatedly fourteen galaxies in two different masks. Similarly, in the A1307 
field, we acquired double spectra for eight targets. These double-redshift measurements allows us to
obtain information about realistic errors (including systematic ones) relative to the redshift estimate 
procedure. We find a no significant difference, with values of $46 \pm 185$ km s$^{-1}$ for spectra in the A76 
field, and $42 \pm 200$ km s$^{-1}$ in the field of A1307. In addition, comparing our redshift estimates
with that retrieved from NED and SDSS database, we find a negligible mean difference of
$7 +\pm 240$ km s$^{-1}$.

\begin{figure*}
\centering
\includegraphics[width=\textwidth]{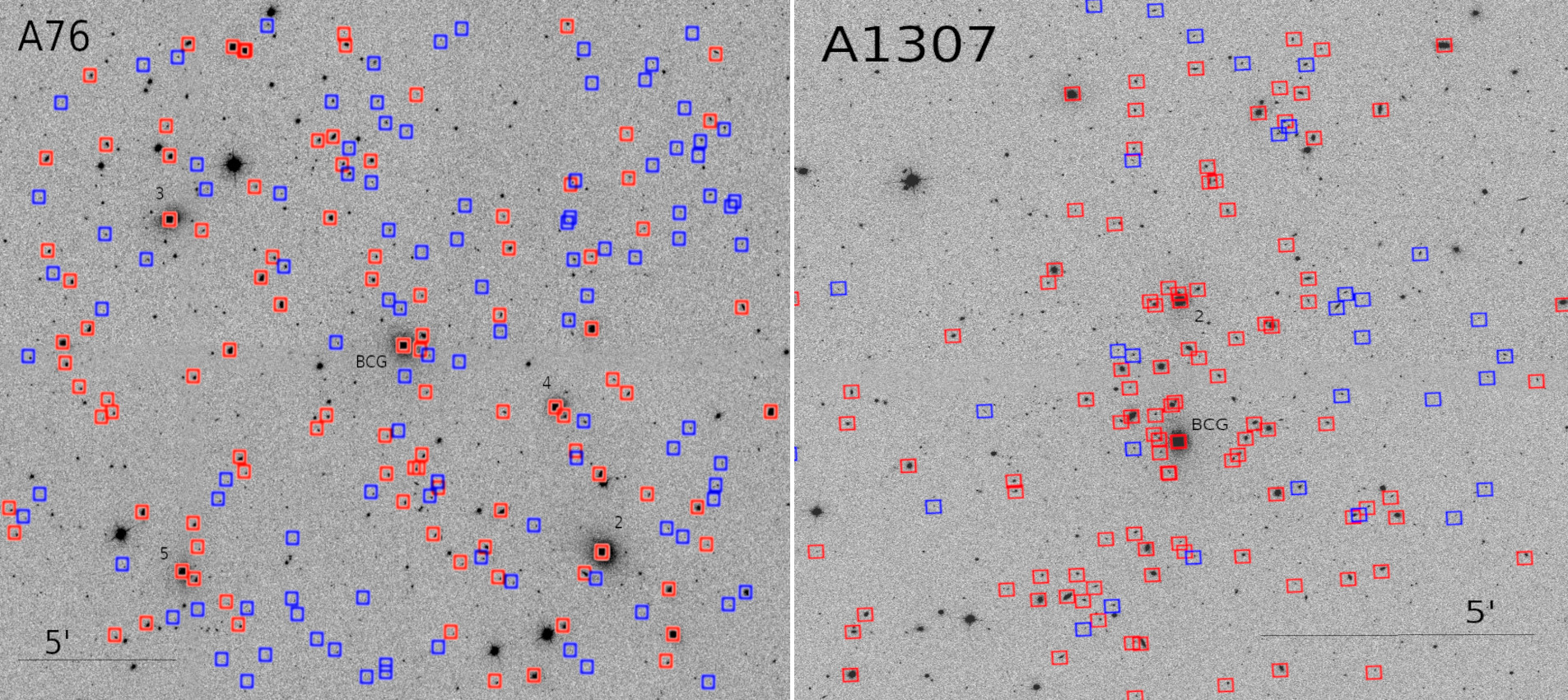}
\caption{Spectroscopic sample of galaxy targets considered in this work in the A76 and A1306 fields, 
superimposed to SDSS r-band images. Red and blue squares correspond to the cluster member and non-member 
galaxies, respectively. The numbers mark the brightest galaxies as explained in the text.}
\label{fig:redshifts}
\end{figure*}

The spectroscopic sample is not fully complete. Comparing the photometric sample of likely galaxy 
members (see Sect. \ref{sec:spatial_distrib}) with the spectroscopic one, we find that the completeness in 
the A76 field is about 40\% in the central region of the cluster and decreases down to 10\% for radii 
$r \sim r_{200}$. Similarly, the spectroscopic completeness in A1307 is about 65\% in the inner region, 
while this decreases down to 10\% for radii $r \sim r_{200}$. 

Tables\footnote{Tables \ref{tab:catalog_A76} and \ref{tab:catalog_A1307} are available in electronic form at 
the CDS via anonymous ftp to cdsarc.u-strasbg.fr (130.79.128.5) or via \url{http://cdsweb.u-strasbg.fr/cgi-bin/qcat?J/A+A/}} 
\ref{tab:catalog_A76} and \ref{tab:catalog_A1307} list the spectroscopic redshifts considered in
this work. Column 1 lists the ID number, whith cluster members marked with an asterisk. Cols. 2 and 3 include 
the equatorial coordinates of galaxies, RA and Dec in J2000 system; Col. 4, the heliocentric radial 
velocity ($v=cz$) and their errors ($\Delta\textrm{v}$); Cols. 5 and 6, the complementary $g'$ and $r'$ 
$dered$\footnote{$dered$ magnitudes are the extinction-corrected model ones, provided by SDSS archive, which 
considers the \citet{Sch98} maps for reddening corrections.} magnitudes obtained from SDSS DR16 photometric 
database; and a last column with some comments regarding particular features of some galaxies, such as the 
presence of 'emission line galaxies' (ELGs). Figure \ref{fig:redshifts} shows the spatial distributions of 
these spectroscopic samples in the A76 and 1307 fields.

Since our spectroscopic samples are not complete, we also consider the publicly available photometry from 
SDSS DR12 archive, so complementing our redshift survey. We work with $dered$ magnitudes 
$g'$ and $r'$ retrieved within the same regions exposed in Sect. \ref{sec:spectroscopy}. In agreement with
SDSS DR12 estimates\footnote{see \url{https://www.sdss.org/dr12/imaging/other\_info/}}, we find that the mean
depth (at $\sim$90\% completeness) of this photometry is $r'=21.5$.

\begin{figure} 
\centering
\includegraphics[width=\columnwidth]{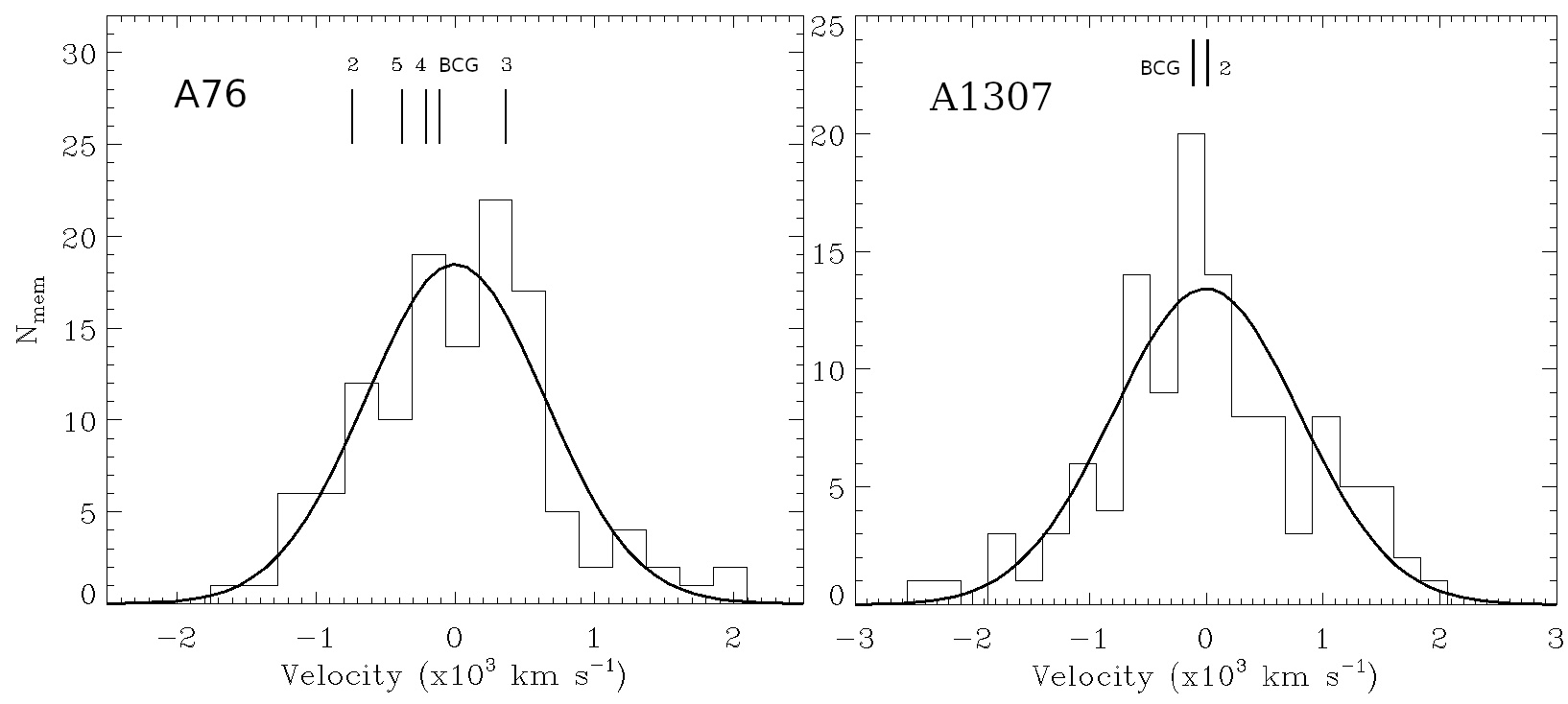}
\caption{Velocity histograms of galaxy members in the cluster rest frame. The Gaussian curves represent the 
reconstruction of the velocity distribution according to the biweight method. The velocity corresponding to 
the BCG, and other luminous galaxies are labelled as 'BCG', '2', '3', '4' and '5'.}
\label{fig:histogram}
\end{figure}

\subsection{X-ray data}
\label{sec:x-ray}

The {\it XMM-Newton} observations are publicly available in the archive, with IDs 0800760501 and 0827010801 
for A76 and A1307 clusters, respectively. The A76 cluster was recently observed with an exposure time of 
12~ks (Obs. ID: 0920010401; PI: L. Lovisari) as part of a follow-up of an X-ray selected sample of groups, while 
A1307 was observed during 27~ks as part of CHEX-MATE Cluster Heritage project (PIs: M. Arnaud and S. Ettori; 
\citealt{Arn21}). Both acquisitions were carried out using the EPIC MOS camera in the 0.5 to 2 keV band. We 
reduce the X-ray images and spectroscopic data using the SAS v20.0 software closely following the technique 
described in~\citet{chon15}. We remove hot pixels and subtract the background. The final images obtained are 
shown in Fig.~\ref{fig:A76A1307_images}.

\begin{figure*}[ht!]
\centering
\includegraphics[width=\textwidth]{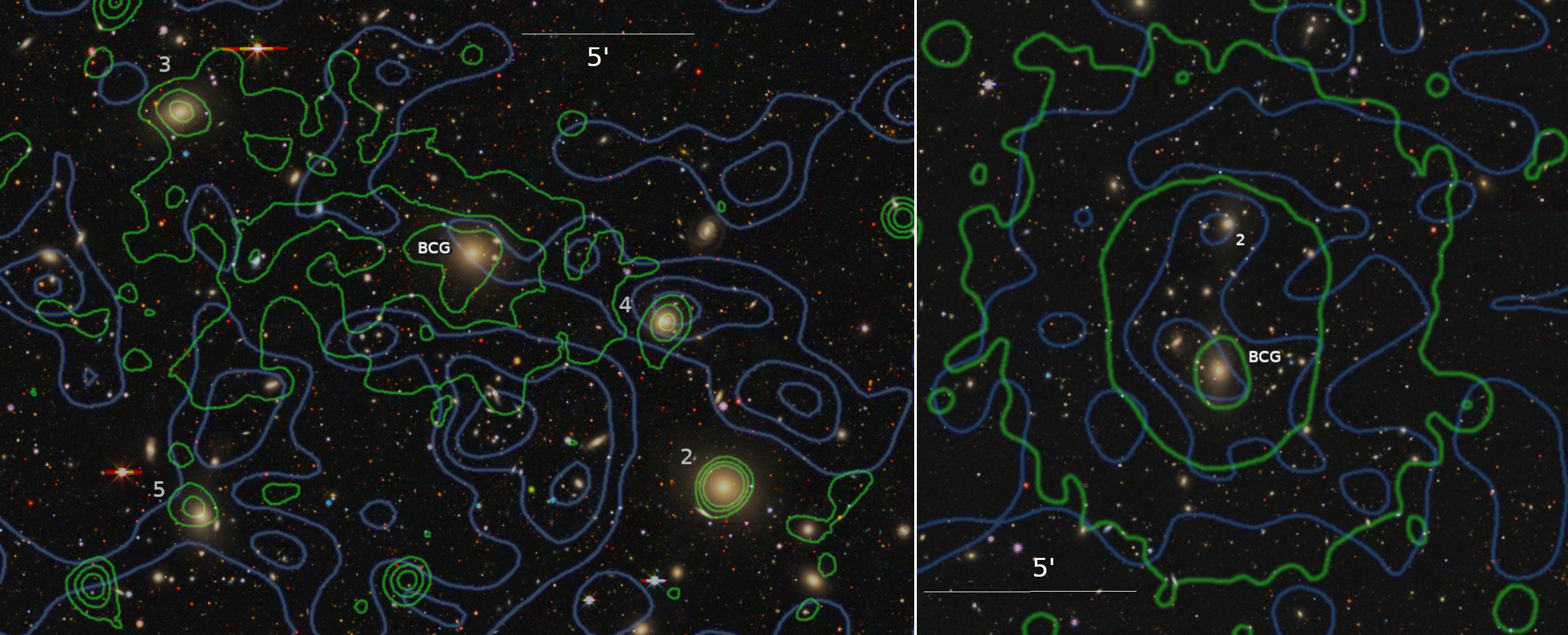}
\caption{Colour image obtained from the DR9 Legacy Survey of A76 (left) and A1307 (right). Superimposed, we
plot contours corresponding to the X-ray surface brightness (green) and the isodensity galaxy distribution 
of likely cluster members (grey-blue). The BCG, "2", "3", "4" and "5" labels indicate the central bright cluster 
galaxy, and the other luminous galaxies. North is up and east to the left in both panels.}
\label{fig:contours}
\end{figure*}
 
A76 has a quite elongated shape along the east-west direction, and its brightness is very low, shallow and 
very clumpy. We find that the coordinates of the position of the global, diffuse  X-ray emission, 
obtained by fitting five ellipses of semi-major axes from 4 to 8 arcmin, vary within a region of about 
1.2 arcmin. With this large centre shift, we can conclude that the X-ray diffuse emission is very irregular 
and its centre is not well defined (see e.g. \citealt{Boh10,Cam22}). However, the brightest region is placed 
to the west part of the cluster, while the X-ray emission extends, like a tail, to the east. 

A1307 presents, however, a much more compact and bright emission, with a single peak in the south which seems to 
be embedded in a larger south-north tail. The X-ray emissions of A76 and A1307 are about 0.7 and 1.3 Mpc large, 
respectively, along their most elongated directions. \citet{Cam22} performed a morphological analysis of A1307 
together with a sample of CHEX-MATE clusters. With a concentration parameter of $c = 0.27$ and a centre shift 
parameter of $w = 0.038$, A1307 is found as one of the most disturbed clusters in the sample (see also Fig. 1 
in \citealt{Lov24}). Furthermore, both clusters show a high central entropy, a very extreme value of 400 keV 
cm$^2$ in the case of A76 (OTA13) and $136 \pm 25$ keV cm$^2$ for A1307. This also characterises the clusters 
as disturbed.

\begin{figure} 
\centering
\includegraphics[width=\columnwidth]{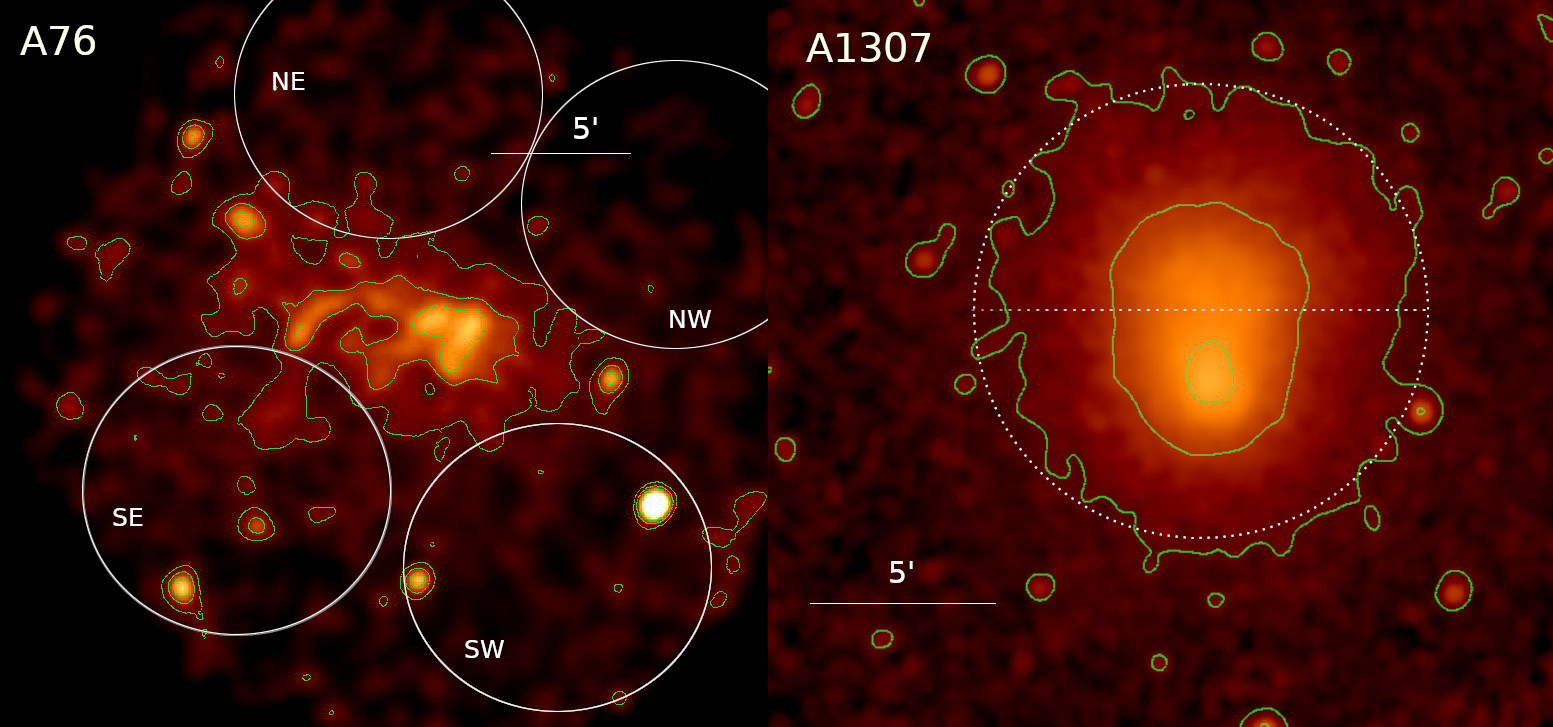}
\caption{{\it XMM-Newton} images of the clusters A76 and A1307 in the 0.5 to 2 keV energy band.
The images have been smoothed with a 7 pixels box size in order to reduce the noise. Green contours correspond 
to 8, 15 and 20-$\sigma$ detection levels in A76 panel, while A1307 contours match with 6, 30 and 
90-$\sigma$ detections. The circle SE in the A76 frame indicates the subcluster at the south-east detected 
by OTA13 in the {\it XMM-Newton} observations. We perform a comparative analysis of the diffuse signal 
from regions labelled as SE, SW, NE and NW in Sect. \ref{sec:Xprop}. Dotted lines in the A1307 frame demarcate
the two hemispheric regions, north and south, analysed in this work. In both frames, north is up and east is 
to the left.}
\label{fig:A76A1307_images}
\end{figure}

For the spectral analysis we treated the sky background as composed by three main contributions when fitting 
the data using the XSPEC tool: the unresolved point sources, the Local Hot Bubble and a cool absorbed thermal 
model when fitting an APEC cluster model to the spectroscopic data.

\section{Optical properties of A76 and A1307}
\label{sec:optical_ana}

\subsection{Member selection and global properties}
\label{sec:members_global}

We analyse the kinematics of A76 and A1307 on the basis of the spectroscopic sample. The first step is
to perform a good selection of member galaxies. One way to select cluster members is from the projected 
phase-space distribution of member galaxies \citep{Woj13} under the assumption of dynamical equilibrium 
\citep{Car97}. This method allows us to identify cluster members from background and foreground
galaxies. Nevertheless, this technique only offers guarantees for large spectroscopic samples, including 
typically 200 galaxies or even more. We apply this method to our sample, but we do not find reliable 
results. Consequently, in our study, we employed a similar and simplified technique based on the projected 
($r$, $cz$) space. Our approach involves an iterative 2.7$\sigma_\textrm{v}$ clipping in $cz$,
assuming the radial profile of the expected velocity dispersion \citep{Mam10}. Initially, we determine 
the mean velocity of the cluster and compute a first velocity dispersion using the root mean square 
($rms$) estimator. Through successive steps, we arrive at stable and converging values for $\bar{\textrm{v}}$
(mean velocity) and $\sigma_\textrm{v}$ (velocity dispersion). This method yield a selection of 122 and
116 cluster members for A76 and A1307, respectively. So, discarding 2 and 3 foreground galaxies and 
107 and 44 background galaxies in each field, respectively. Our cluster member selection compiles
galaxies within the range 0.0337<z<0.0463 for A76 and 0.07<z<0.09 for A1307. Figure \ref{fig:histogram}
shows the velocity histograms obtained from this cluster selection and the Gaussian velocity reconstruction 
according to the bi-weight method.

With the aim in mind of estimating a robust mean velocity and velocity dispersion, we apply two different methods, 
the root mean square ($rms$) and the bi-weight scale estimator \citep{Bee90}. This last procedure provides 
more solid estimates with respect to the former for samples showing possible inhomogeneities. Applying the bi-weight
to our redshift samples of cluster members we obtain mean velocities $\bar{\textrm{v}}=11909 \pm 69$ km s$^{-1}$ 
and $24450 \pm 94$ km s$^{-1}$, and velocity dispersion $\sigma_\textrm{v}=650_{-64}^{+74}$ and $863_{-87}^{+101}$ 
km s$^{-1}$, for A76 and A1307, respectively. On the other hand, the $rms$ estimate of the velocity dispersion 
yields a values of $669 \pm 55$ and $863 \pm 42$ km s$^{-1}$ for A76 and A1307, respectively. Table 
\ref{tab:properties} compiles these findings. The good agreement between the two $\sigma_\textrm{v}$ estimates, even for 
the two clusters, makes evidence that there is no strong hints for inhomogeneities in the velocity distribution. 
However, in order to check possible deviations of $\sigma_\textrm{v}$ from the mean along the cluster, we analyse 
in Sect. \ref{sec:velocity} the stability of this magnitude with the distance to the cluster centre (assumed as 
the BCG position).

\begin{table}[ht]
\begin{threeparttable}
\caption{Positions and global properties of A76 and A1307.}             
\label{tab:properties}
\begin{center}
\begin{tabular}{lcc}
\hline
\hline
 & {\rm A76} & {\rm A1307} \\
\hline
% RA,Dec ($hh\! : \! mm \! : \! ss.s$; $^\circ \! :\! ^\prime \! : \! ^\prime$$^\prime$ ; J2000) & 00:39:55.94 \enspace +06:50:55.5 &  11:32:51.17 \enspace +14:27:40.3 \\
RA  ($hh\! : \! mm \! : \! ss.s$) &  00:39:55.94 &  11:32:51.17 \\
Dec ($^\circ \! :\! ^\prime \! : \! ^\prime$$^\prime$) & +06:50:55.5  & +14:27:40.3  \\
N$_{gal}$ & 122                 & 116 \\
$\bar{\textrm{v}}$ (km s$^{-1}$)          & 11909$\pm$69        & 24450$\pm$94 \\
$\sigma_\textrm{v}$ (km s$^{-1}$)         & $650_{-64}^{+74}$   & $863_{-87}^{+101}$  \\
$r_{200}$ ($h_{70}^{-1}$Mpc)              & 1.2$\pm$0.2         & 1.5$\pm$0.2  \\
M$_{200,\rm{dyn}}$ ($\cdot10^{14}$ M$_{\odot}$)  & $1.8 \pm 0.6$       & $4.3 \pm 1.2$ \\
M$_{200,\rm{MAM}}$ ($\cdot10^{14}$ M$_{\odot}$)  & $2.9^{+1.6}_{-0.6}$ & $7.7^{+2.8}_{-3.0}$ \\
M$_{500,\rm{dyn}}$ ($\cdot10^{14}$ M$_{\odot}$)  & $1.1 \pm 0.3$       & $2.7 \pm 0.7$  \\
M$_{500,\rm{MAM}}$ ($\cdot10^{14}$ M$_{\odot}$)  & $2.1^{+0.5}_{-0.4}$ & $4.5^{+1.1}_{-1.0}$ \\
M$_{500,\rm{X}}$ ($\cdot10^{14}$ M$_{\odot}$)    & $1.9 \pm 0.7$          & $2.75 \pm 0.4$ \\
M$_{500,\rm{SZ}}$ ($\cdot10^{14}$ M$_{\odot}$)   & $1.7 \pm 0.2$       & $3.9 \pm 0.2$ \\
T$_{\rm{X}}$ (keV)                        & $3.3 \pm 0.1$          & $\sim 4.3 \pm 0.4$  \\
\hline                 
\end{tabular}
\begin{tablenotes}
\item \footnotesize{\textbf{Note:} the error on M$_{500,\rm{X}}$ is mainly the uncertainty of the X-ray 
luminosity mass relation. The T$_{\rm{X}}$ of the A76 cluster corresponds to the value provided by 
\citet{Ota13}. For more precise values on the temperature see the temperature profile of A76 in this 
paper. For precise values of the temperature profile of A1307 see the profile in Fig.~8. The SZ mass, 
M$_{500,\rm{SZ}}$, has been retrieved from PSZ2 {\it Planck} catalogue.}
\end{tablenotes}
\end{center}
\end{threeparttable}
\end{table}

A76 presents a bright cluster galaxy (BCG; ID 122), well centred with the X-ray central diffuse emission 
and coincident with a moderate central, a fact that OTA13 also observe using the X-ray {\it Suzaku} data. 
However, this bright galaxy, despite of showing a normal BCG properties, is not the brightest cluster member 
in optical bands. We find three other galaxy members, the IDs 54, 197 and 79 labelled as "2", "3" and "4" 
in Fig. \ref{fig:contours}, even brighter than the BCG. Moreover, we identify a fifth galaxy member, the 
ID 193, labelled with a "5" in Fig. \ref{fig:contours}, with a similar magnitude to the BCG. All these four 
galaxies (the "2", "3", "4" and "5") present a more intense X-ray emission than that of the BCG, and show 
clear shifts with respect to the centre of the cluster, larger than 240 kpc in projected distance. None 
of these BCGs present emission lines in their optical spectra, so they are passive early-type galaxies 
showing no star formation or AGN activity. The X-ray halos around the BCGs show some extent, only the halo 
of BCG~4 is not resolved and could be a point source. The velocity distribution of these 
galaxies is also shown in Fig. \ref{fig:histogram}. The BCG is quite well centred in the velocity distribution 
and the other brightest galaxies present large velocity offsets, even larger than 200 km s$^{-1}$. In summary, 
A76 presents a large accumulation of bright galaxies: a BCG, quite well centred in the velocity field and in 
the X-ray diffuse emission, and several other bright galaxies, outside the cluster core showing large velocity 
offsets.

A1307 also presents a clear BCG (the ID 84), coincident with the X-ray maximum emission of the cluster. In 
addition, there is a second bright galaxy (ID 79) placed at 3.2 arcmin (about 300 kpc in projected distance) 
towards the north. Therefore, the BCG and the galaxy "2" follow a south-north alignment, in agreement with the 
south-north tail observed in the X-ray surface map. The radial velocity of the galaxy "2" is, in practice, 
the same as the mean cluster velocity, as it is shown in Fig. \ref{fig:histogram}. 

We detect 78 and 32 galaxies that present emission lines, mainly [OII], in the A76 and A1307 fields, respectively. 
They are conveniently labelled as "ELG" in tables and \ref{tab:catalog_A76} and \ref{tab:catalog_A1307}.
The S/N and resolution of our spectra allow us to detect [OII] lines as far as they
show equivalent width >10 \AA. The presence of emission lines in the spectra is related to star-forming
processes in galaxies. In A76, we find that 13 ELGs are cluster members, while in A1307, only four members
present emission lines. This means that A76 shows a population with about 10\% of galaxy members showing star 
formation, while in A1307, the star-forming galaxies only represent a fraction as low as 3.4\%. The
high galaxy density environments and hot intra-cluster medium (high T$_X$) in clusters produce strong 
quenching effects in their galaxy population \citep{Lag08}. Therefore, A76 shows a typical low fraction of
star-forming galaxies. However, this fraction is especially low in A1307, so we can infer that the whole 
galaxy population with star formation has been dramatically quenched in this system. 

\subsection{Velocity field}
\label{sec:velocity}

In general, deviations of the global velocity distribution with respect to the Gaussian profile are usually 
good indicators that the internal dynamics of clusters is not relaxed, or may even warn of the presence of 
substructure (see e.g. \citealt{deCar17}; \citealt{Rib11}). In this sense, the skewness and kurtosis are 
appropriate measures to evaluate the departure from gaussianty. The skewness is related with the asymmetry 
of the velocity distribution, while the kurtosis indicates distributions showing thinner or fatter 
tails. Our samples present a skewness of $0.05 \pm 0.17$ and $-0.11 \pm 0.16$ in A76 and A1307, respectively. 
Similarly, we find a kurtosis of $-0.07 \pm 0.26$ and $0.10 \pm 0.25$ in both cluster member samples. 
Errors were computed from the standard deviation of the values obtained after performing a Markov chain 
Monte Carlo (MCMC) method with 10000 simulations assuming a Gaussian profile with an average centre equal 
to zero and a standard deviation equal to one, sampled with 122 and 116 points for each cluster sample, 
respectively. So, in our case, the skewness and kurtosis are compatible with zero, which mean that the global 
velocity distribution of A76 and A1307 are quite symmetric, following a quite regular Gaussian profile, 
which suggests that both clusters are not dramatically disturbed and with no clear evidence of the presence 
of substructure.

\begin{figure} 
\centering
\includegraphics[width=\columnwidth,height=7cm]{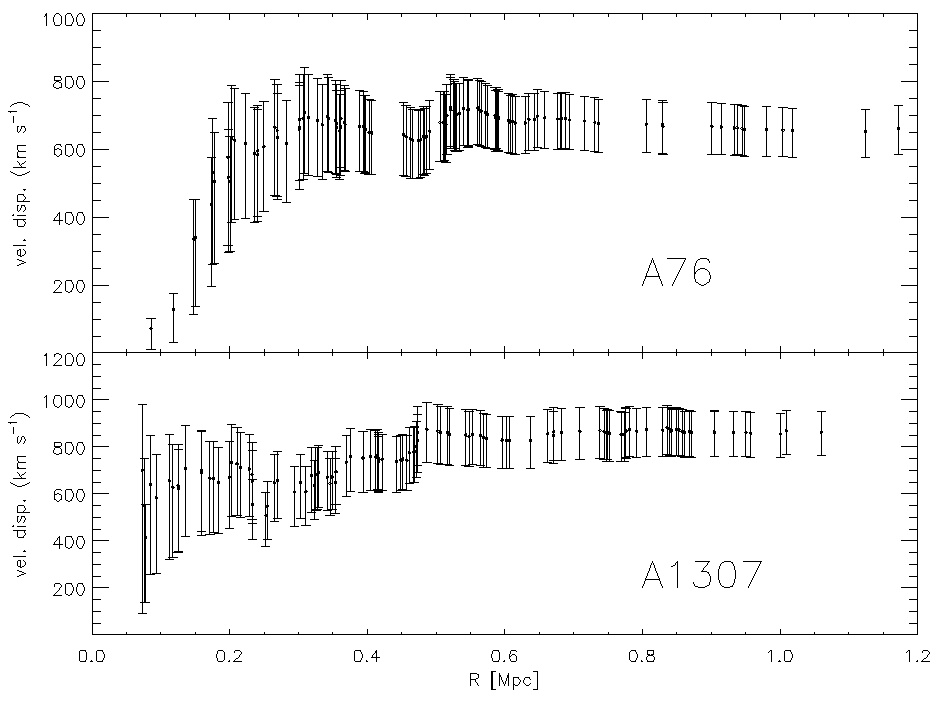}
\caption{Variation of the line-of-sight velocity dispersion, in the cluster rest frame, with the 
cluster-centric distance. The cluster centre is assumed to be the BCG position. The velocity dispersion 
values are computed using the bi-weight estimator and considering all galaxy members enclosed in that radius. 
The first value is calculated from the first five galaxies closest to the centre. The error bars are at the 
68\% confidence level.}
\label{fig:vel_disp_r}
\end{figure}

Figure \ref{fig:vel_disp_r} shows the integral $\sigma_\textrm{v}$ variation as radial profile for A76
and A1307, respectively. In the case of A76, the $\sigma_\textrm{v}$ profile remains stable beyond 0.2 Mpc.
However, it is worth noting the pronounced drop in values in the innermost part of the cluster, where the
$\sigma_\textrm{v}$ takes values even lower than 200 km s$^{-1}$, probably typical values of the internal 
velocity dispersion of the BCG. The profile of the $\sigma_\textrm{v}$ variation in A1307 is completely 
different. The $\sigma_\textrm{v}$ in A1307 is quite flat over the whole cluster but it shows a
differential behaviour in the inner part with respect to the external region. In fact, when we compute 
the $\sigma_\textrm{v}$ considering only members at radii $r<0.47$ Mpc, we find $745 \pm 118$ km s$^{-1}$. 
On the contrary, the $\sigma_\textrm{v}$ for cluster members in the external region $r>0.47$ Mpc is 
$827 \pm 81$ km s$^{-1}$. This small difference, at 1-$\sigma$ level discrepancy, between inner and outer 
velocity dispersion may suggest that the dynamics of the cluster could be different in the inner part with
respect to the regions at $r \gsim 0.5$ Mpc. We analyse this in detail in following sections.

Another parameter commonly used to establish whether a galaxy cluster may present a disturbed dynamical state 
is the offset between the BCG peculiar velocity and the average cluster velocity. However, as it is shown in
Fig. \ref{fig:histogram}, in both A76 and A1307 the BCG velocity offset is very small, of about 110 km 
s$^{-1}$ in both cluster, which indicates that both clusters do not present significant signs of unrelaxation 
in this respect, which agrees with the fact that the global velocity distributions present likely Gaussian 
shapes.

\subsection{Two-dimensional galaxy distribution}
\label{sec:spatial_distrib}

In practice, spectroscopic observations cover the galaxy population of the whole clusters partly, not only
due to the spatial incompleteness of the sample, but also because the spectroscopic observations miss the 
faintest galaxies. So, in order to get information of the distribution of a more complete galaxy member 
sample, we use the SDSS photometric database, data release 16 (DR16). Using the $g'$ and $r'$ $dered$ 
magnitudes (which are corrected by reddening), we build the ($g'-r'$ vs $r'$) colour--magnitude diagramme (CMD). 
This technique allows us to obtain likely galaxy members by selecting galaxies within the red sequence (RS)
locus. From this diagrammes we measure RS of $g'-r'=-0.028(\pm0.004) \times r'+1.37(\pm0.07)$ and 
$g'-r'=-0.031(\pm0.001) \times r'+1.31(\pm 0.02)$ for A76 and A1307, respectively. We select likely
early-type members as galaxies showing colours with upper limit $g'-r'<$RS$+3 \cdot rms$ and late-type 
galaxies with $g'-r'>-0.054 \cdot r'+1.212$ for the A76 field. Similarly, for A1307, we select likely 
members that galaxies showing $g'-r'<$RS$+3 \cdot rms$ and $g'-r'>-0.1509 \cdot r'+3.013$. In both cases, 
we impose the restriction $r'<22.5$, which is the magnitude limit of the photometric sample. This
selection criteria provide 2105 likely members in a region of $30^\prime \times 50^\prime$ in the A76 field 
and 1109 likely members in a circular region of 12$^\prime$ radius in the A1307 field.

We use these photometric samples in order to investigate the galaxy distribution of A76 and A1307. With 
this aim in mind, we construct the isodensity galaxy maps shown in Fig. \ref{fig:contours} (grey-blue 
contours). These maps were obtained by quantifying the contribution of 2105 and 1109 small Gaussians with 
$\sigma=1$ arcsec width, placed at the positions of the likely members, over a grid of $200 \times 200$ 
points. The contour maps obtained reveal two very different configurations. While A76 shows a very
spread galaxy density, that of A1307 is quite compact. The 2D galaxy distribution of A76 shows several
peaks, but none of them are centred on the BCG. We detect one galaxy concentration 5 arcmin to the 
south of the BCG and a secondary overdensity around the BCG-4, but slightly shiftted to the south-west.
In fact, none of the five brightest galaxies shows high galaxy density around, with the exception of
the BCG-4, that shows a quite irregular and elongated galaxy overdensity around it. The case of A1307
is much more regular. The galaxy distribution of A1307 reveals a double-peaked distribution clearly 
elongated in the south-north direction. The highest galaxy density is around the main BCG, while the
secondary peak is well centred on the BGC-2 position. 

From the analysis above, we can conclude that the galaxy distribution of A76 is very shallow
while that of A1307 is compact and elongated in the south-north direction. This means that both A76
and A1307 present a galaxy distributions and X-ray surface brightness profiles that are almost coincident.
As we have shown in Sect. \ref{sec:x-ray}, The X-ray emission of A76 is quite weak, shallow and clumpy, in 
agreement with its 2D galaxy density distribution. Similarly, A1307 presents a compact X-ray emission, with
a peak centred in the BCG and elongated in the south-north direction. So, the X-ray configuration perfectly 
match the galaxy component of A1307.

\subsection{Analyse of substructures}
\label{sec:spa_vel_correlat}

The 1D (velocity) analyses reveals that velocity distributions of galaxy members in A76 and A1307 
follow an almost Gaussian profiles, which is confirmed by the values obtained for skewness and kurtosis (see Sect. 
\ref{sec:velocity}). Moreover, to provide an additional assessment of the dynamical state of the clusters,
we compute the Anderson-Darling coefficient A$^2$ \citep{Anderson54}, which measures departures from 
Gaussianity of the line-of-sight velocity distribution. As described in \cite{Pizzuti20}, A$^2$ is a good 
indicator of the systematic effects induced by lack of dynamical relaxation. We found A$^2=0.38$ for Abell 
76 and A$^2=0.56$ for Abell 1307, which suggests that the former could be in a more relaxed dynamical state. 
However, the significance level is large for both clusters, $\alpha > 0.15$, indicating that the null 
hypothesis of an underlying Gaussian distribution cannot be rejected at high confidence. Therefore, additional 
assessments are needed to better investigate the morphological properties of these clusters. 

\begin{figure} 
\centering
\includegraphics[width=\columnwidth,height=4cm]{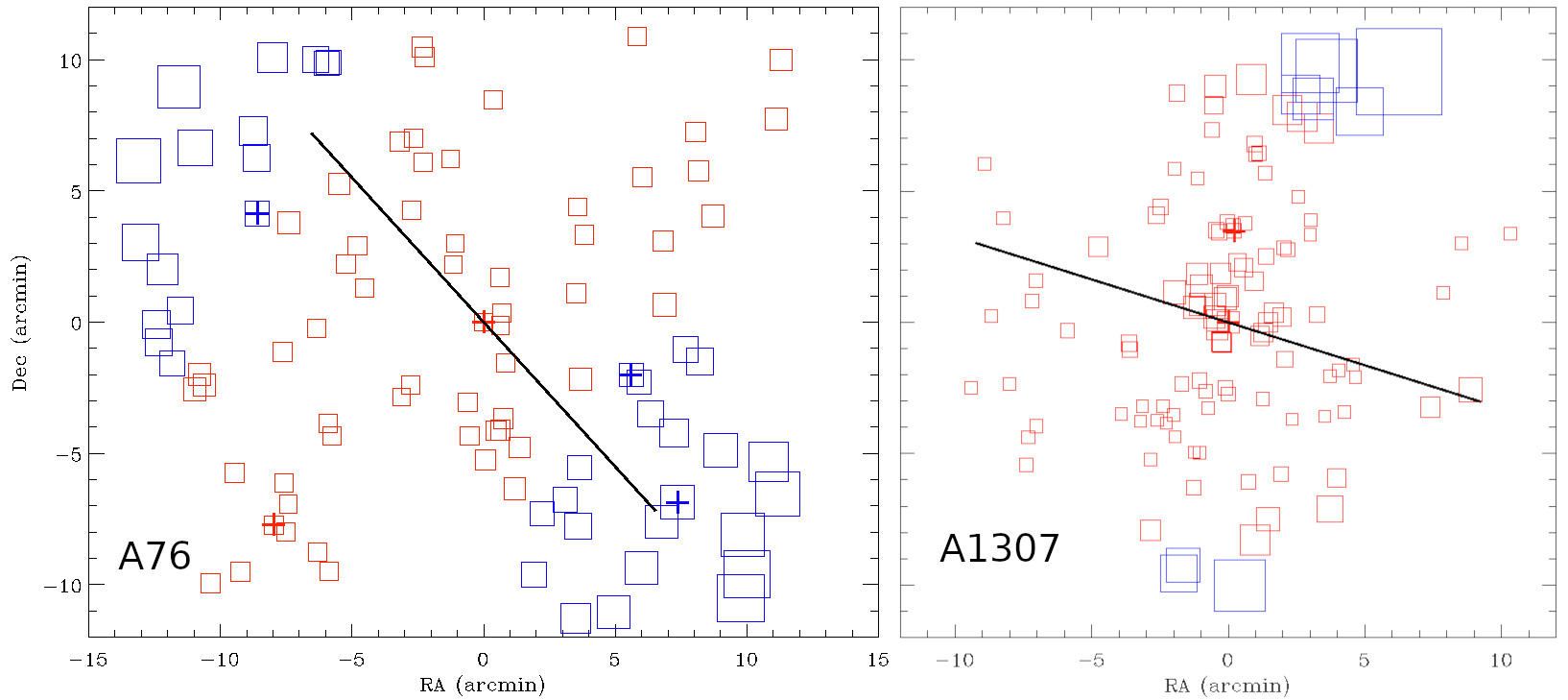}
\caption{Spatial distribution of the cluster members, marked with squares and a size proportional to 
$exp(\delta_i)$ computed using the $\delta_i$ deviations obtained with the DS test. Red and blue correspond 
respectively to galaxies with lower and higher $\delta_i$ deviations from the mean, $<\delta_i>$. The 
large black lines represent the directions of the velocity gradients in each cluster region.
Crosses indicate the BCGs positions in each cluster.}
\label{fig:DS_delta}
\end{figure}

In order to check the presence of galaxy clumps with their own kinematical
entity, we apply several other tests in the 3D (spatial and velocity) space. In the past decades, numerous 
techniques have been proposed to determine the presence of substructures within clusters. These procedures 
are usually based on the positions and velocities (3D space) of individual galaxy members, so 
analysing possible space-velocity segregation. To this aim, one of the most used algorithm is the 
\citeauthor{Dress88} (DS) test \citep{Dress88}, which search for galaxy clumps whose mean velocities 
and/or dispersion deviate from the global cluster values. We apply this procedure to our samples by
running 1000 Monte Carlo simulations. Figure \ref{fig:DS_delta} shows the $\delta$-statistics test over 
the inner and best sampled region. For both A76 and A1307 clusters, we obtain similar results. The clusters
have clear main bodies that dominate their respective central regions (red squares in Fig. 
\ref{fig:DS_delta}) with several galaxies in their outskirts showing larger deviations (blue squares in 
Fig. \ref{fig:DS_delta}). A76 shows deviations towards the north-east and south-west, almost coincident 
with the elongation present in the X-ray surface brightness map. Similarly, A1307 shows some galaxy 
members with not negligible deviations in the north and south edge of the cluster, also following the 
direction of the X-ray emission profile. However, in the case of A1307, the low number of members with
high deviations in the external zone of the cluster suggests that these few galaxies present decoupled 
kinematics that are now slowly falling onto the cluster core. We find a mean deviation $<\delta_i>=4.09$ 
and 6.3 for A76 and A1307, respectively, a cumulative deviations of $\sum_i \delta_i=41.6$ and 726.5. 
The likelihood ratio test statistic is 30.14\% and 57.03\%, with a p-values of 0.036 and 0.00,
for this statistics applied to the A76 and A1307 samples, respectively. In summary, the results 
here obtained put on the table the possibility that A76 presents a heterogeneous kinematics characterised 
by a central main body, which includes the BCG and BCG-5 with galaxies around them, and external 
clumps, slightly decoupled from the main kinematics of the cluster, that includes galaxy clumps dominated 
by the BCG-2, BCG-3 and BCG-4. On the contrary, A1307 presents a more homogeneous kinematics made up of 
a single galaxy clump, which are dominated by the BCG and the BCG-2.

We perform a second test in order to investigate possible space--velocity correlations. This test 
consists of analysing the presence of velocity gradients. With this purpose in mind we fit a plane to the 
full cluster member samples. For A76, we obtain the expression $\Delta\textrm{v}$=-7.95x+8.77y-38.88, and
for A1307 we obtain $\Delta\textrm{v}$=-22.50x+7.33y-4.68; where x and y correspond to the positions of
the galaxies following the R.A. and Dec. coordinates (positive values are taken towards the north and west 
directions), in arcmin, with respect to the BCG. The plane corresponding to the A76 cluster sample shows
a maximum gradient of 65($\pm 184$) km s$^{-1}$ Mpc$^{-1}$ in a $42^\circ(\pm 4^\circ)$ angle 
(anticlockwise, from north to west; $\Delta\textrm{v}$ takes positive values towards the north-east). 
Similarly, the best plane fit obtained for A1307 presents a maximum gradient of 260($\pm 310$) km s$^{-1}$ 
Mpc$^{-1}$ in a $72^\circ(\pm 8^\circ)$ angle, which means that $\Delta\textrm{v}$ takes higher values
towards the west of the cluster. Therefore, as it is shown in Fig. \ref{fig:DS_delta}, despite that the velocity
gradients do not take very significant values (errors are large in comparison with the gradient obtained), 
we observe that A76 presents a maximum velocity gradient almost coincident with the elongation of X-ray 
surface brightness. On the contrary, A1307 shows a maximum gradient in the ENE-WSW direction, so almost 
perpendicular to the X-ray surface brightness profile. 

A third test (also in the 3D space, that is, positions and velocity) has been further performed in order 
to identify possible substructures. We use a 3D version of the Kaye's  
Mixture Model (KMM) algorithm \citep{Ash94} to associate individual galaxies to each substructure.
This procedure provides a probability that a given galaxy belongs to different galaxy clumps. The KMM
needs an input configuration, so we run the algorithm in two different ways. Given that we have not been able 
to identify any subcluster, first we use 100 lists of galaxies randomly associated to two clumps, and we 
run the algorithm 100 times. This method does not provide consistent results, neither for A76 nor for 
A1307. The results obtained are not stable and we only obtain divergent solutions, without any reliable 
identification of substructures. Then, we run KMM in a second way, by starting with a single input list 
of galaxies including two separate and well defined clumps: one containing galaxies showing high $\delta_i$ 
deviations and a second clump with galaxies showing low $\delta_i$. That is, two input populations defined 
by blue and red members in Fig. \ref{fig:DS_delta}. Again, the result obtained is not reliable, because the 
output probabilities assigned to the galaxies with high $\delta_i$ are even lower than 90\%, which are
not acceptable values (high confidence values would be output probabilities higher than 95-98\%). Thus, the 
KMM test does not provide conclusive proofs of the existence of substructure in either A76 or A1307. 

\begin{figure}
        \centering
        \includegraphics[width=0.8\columnwidth]{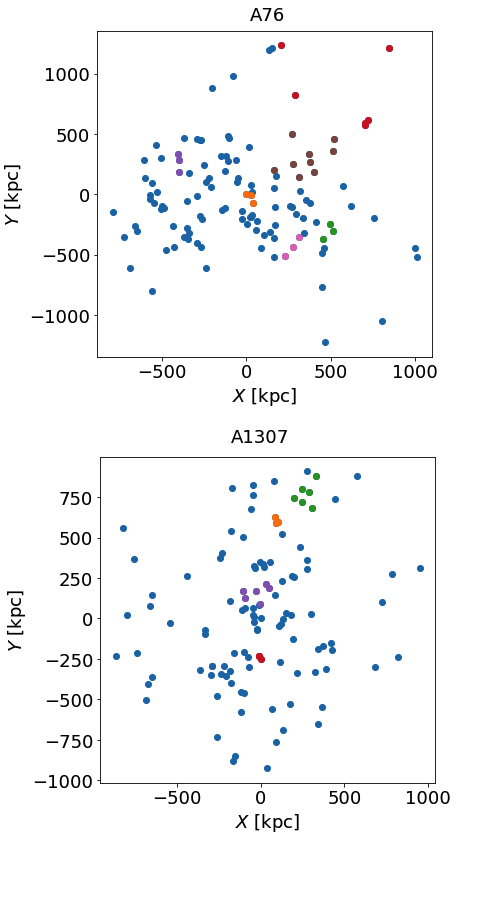}
        \caption{Projected positions of the cluster members (blue) in A76 (top) and A1307 (bottom) fields, 
        where we have marked using other different colours the substructures identified by the DS+ algorithm.}
        \label{fig:DSp}
\end{figure}

Finally, we employed the DS+ algorithm \citep{Benavides23}, an extension of the DS method based on the 
projected phase space of position and line-of-sight velocities, which identifies substructures and estimates 
the probability of each galaxy to belong to a given sub-group. As for the DS case, we run 1000 Monte Carlo 
samplings and we consider as reliable substructures those for which the estimated p-value is $\lesssim 0.01$.
This test revealed that only 22\% of the galaxies of A76 may configure six groups within the cluster, four 
of which made up of three galaxies, one with six galaxies and one with nine. In the case of A1307, the only 
15\% of members compose four sub-groups, two with three members, and two with six galaxies. These results 
reveal that A76 is mostly configured by small groups that contain very few members, without any important 
entity by themselves. Similarly, A1307 contains even less substructures and even poorer than A76.
In Fig. \ref{fig:DSp} we show the spatial distribution of the member galaxies in both clusters with the 
substructures identified by DS+ marked in various colours.

In summary, the outcome of 1D and 3D tests, the overall structure of A1307 seems to be slightly more 
regular with respect to that of A76. Both clusters show very weak hints of presenting substructure, that 
are very poor and no massive at any time, as also indicated by the analysis of the X-ray features, as 
described in the next Section.

\section{X-ray properties}
\label{sec:Xprop}

A76 has been extensively studied by OTA13. They measured global temperature 
and metal abundance and find a T$_X=$3.3 keV and 0.24 solar, respectively. They also find an extremely high
entropy (about 400 keV cm$^2$) in the core of the cluster, within $r < 0.2r_{200}$, a fact that cannot be easily
attributed by either gravitational heating or preheating. OTA13 examine the {\it Suzaku} observations in very much 
detail and conclude that X-ray morphology is clumped and irregular, and the electron density is extremely 
low (10$^{-4}$-10$^{-3}$) for the observed high temperature. So, they affirm that A76 is in an the early phase 
of formation, and the gas content of A76 is now suffering a delayed relaxation after a recent gas heating due to 
a gravitational potential confinement.

Beside the {\it Suzaku} observations, OTA13 also analysed the X-ray data obtained with the {\it XMM-Newton} 
satellite. From these data, they identify a relevant blob (see Fig. 1b in OTA13) towards the south-east of 
the cluster core (see left panel of Fig. \ref{fig:A76A1307_images}), which they attribute to a likely 
group-scale system that has undergone gas heating in the cluster potential. However, the signal that 
we detect in this region in the {\it XMM-Newton} data is very weak. In fact, comparing the signal in the southern 
regions (SE and SW) with the northern regions (NE and NW) marked in Fig. \ref{fig:A76A1307_images}, we find 
that the northern areas have an average signal at the background level, while the southern regions are at 
$4-$ (SE) and 2.5$-\sigma$ (SW) above the background. In addition, we do not detect  any isolated emission 
peak coming from any clump within the SE region.

Assuming the global temperature provided by OTA13, T$_{\rm{X}}=3.3$ keV, and the temperature-mass 
and temperature-radius relations by \citet{Arn05} (see Table 2 therein), we estimate a total mass of 
M$_{500}=1.9 \pm 0.1 \cdot 10^{14}$ M$_\odot$ within a $r_{500}=0.61 \pm 0.02 $ Mpc. We will use these values 
in Sect. \ref{sec:mass} to contrast them with those obtained from the dynamical studies of A76. However, we 
anticipate here that we obtain a dynamic mass slightly lower than that obtained from the X-ray for relaxed
clusters. Therefore, from this discrepancy we can infer that A76 shows some hints of a non-relaxed dynamical 
state. We will discuss this argument in detail in the following sections.

A1307 shows a clear north-south asymmetry. The surface brightness is steeper in the south and flatter in the
north. Therefore, one may perfectly interpret the morphology of A1307 as that of a symmetric, round cluster, 
which seems undisturbed in the south, but superposed by extended substructure in the north. The centre of the 
undisturbed main component is in this picture the X-ray peak, which is also coincident with the BCG position. 
So, under the X-ray point of view, both A76 and A1307 clusters appear unrelaxed; A1307 in a state of 
a recent minor merger and A76 in an early stage of formation. In addition, we do not detect any evidence of 
shocks and cold fronts in the X-ray images, which could be taken as an indication that we are now witnessing 
a later phase of minor merger events in both clusters. We estimate a global X-ray temperature of about 
4.3 keV, which is a lower value than that obtained by \citet{Lov24}, 4.9 keV. However, we want to remark 
that Lovisari et al. uses a different approach to determine the temperature attempting to correct for 
temperature variations by constructing temperature profiles from two-dimensional temperature maps giving 
specific weights to each pixel and excluding pixels with extreme values.

We obtain X-ray surface brightness profiles after removing the X-ray sky contribution and instrumental 
background. We fit the profiles with $\beta$- and double $\beta$-models. However, we find that single $
\beta$-models provide a good fit to the outer parts of the profiles, from which we determine the gas density 
and gas mass profiles. Only for the southern part of A1307 the double $\beta$-model provides an interesting 
alternative. The fit parameters for the profiles are given in Table \ref{table:a1307}. The magnitudes M$_{500}$ 
and M$_{gas}$ are total and gas mass inside $r_{500}$ radius. The parameter $f_{gas}$ is the gas mass fraction 
of the total mass, $r_{500}$ is the radius projected in the sky, L$_{\rm X}$ is the total X-ray luminosity 
in the [0.5-2.0] keV band, M$_{500}($L$_{\rm X})$ is the mass estimated from the L$_{\rm{X}} - $M relation, 
T$_{\rm{X}}$ is the measured X-ray temperature, also inside $r_{500}$, $r_c$ is the core radius, and $\beta$ 
is the slope parameter of the electron density profile.

\begin{figure}
\centering
\includegraphics[width=\columnwidth,height=6.5cm]{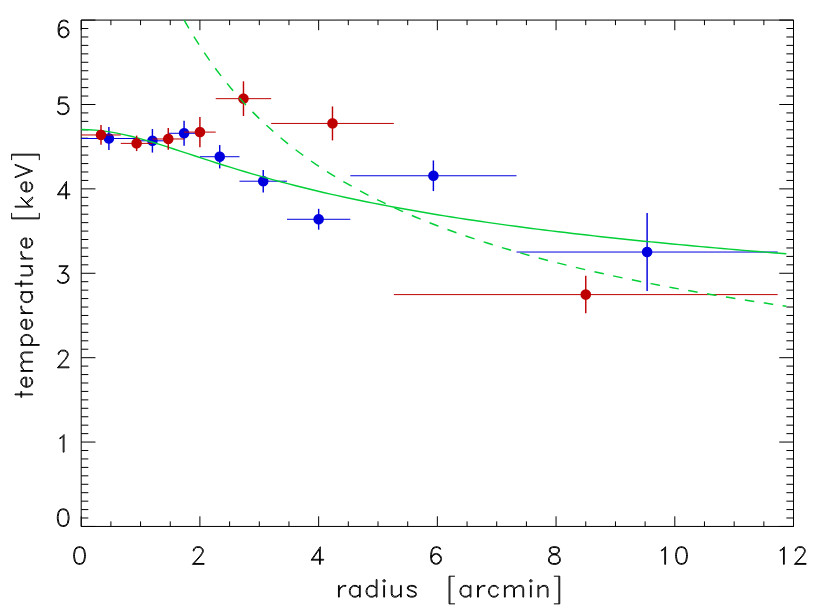}
\caption{Temperature profiles of A1307 fitted by a polytropic model. The temperature profile of the southern 
hemisphere is shown by the red and that of the northern hemisphere by blue data points. The lines show the fits 
of polytropic models: solid line (north) and dashed line (south), where for the latter only the outer part of the 
profile was fitted.}
\label{fig:a1307_tempx}
\end{figure}

The intra-cluster medium (ICM) temperature was determined from the X-ray spectra after subtracting the 
background contribution. We estimate the temperature in eight and seven concentric radial bins in the north and 
south hemispheres, respectively. These bins were constructed in order to have a minimum number of 2000 photons 
per bin. The resulting temperature profile is shown in Fig. \ref{fig:a1307_tempx}. The outer temperature profile 
of the southern region can be approximated by a polytropic model with a $\gamma$ parameter of 1.8. Similarly, 
the northern region is fitted with a $\gamma=1.2$. We also assume the Eq. (4) of \citet{Vik06} to approximate 
the temperature profile. In this formula we discard the part describing the central temperature decrease, which 
allows for a cool-core and not so relevant for A1307. The equation applied thus has the following form:

\begin{equation}
\label{eq:T}
    T(r) = T_0~\frac{(r/r_c)^{-a}} {(1 + (r/r_c)^b)^{c/b}} \ \ \ .
    %~ \frac{(r/r_t) + (T_c/T_0)} {r/r_t + 1}
\end{equation}

\begin{table}
\caption{X-ray properties of A1307.}             
\label{table:a1307}
 \centering          
\begin{center}
\begin{tabular}{lcc}     % 3 columns
\hline
\hline
  & {\rm South} & {\rm North} \\
\hline
M$_{500}$ $(\cdot 10^{14}$ M$_{\odot}$)       & $2.2 \pm 0.3$  & $2.8 \pm 0.4$ \\
M$_{gas}$ $(\cdot 10^{13}$ M$_{\odot}$)       & $3.0 \pm 0.15$ & $4.5 \pm 0.2$ \\
f$_{gas}$                                     & $14 \pm 2 \%$  & $16 \pm 2\%$ \\
r$_{500}$ (arcmin)                            & $9.8 \pm 0.5$  & $10.6 \pm 0.5$ \\
L$_{\rm{X}}$ ($\cdot 10^{44}$ erg s$^{-1}$)   & $0.62 \pm 0.02$ & $0.74 \pm 0.02$ \\
M$_{500}(L_{\rm{X}})$ $(\cdot 10^{14}$ M$_{\odot}$)  & $2.3 \pm 0.05$ & $2.6 \pm 0.05$  \\ 
T$_{\rm X}$ (keV)                             & $4.3 \pm 0.3$     & $4.2 \pm 0.3$ \\
$n_{e0}^{\star}$ ($\cdot 10^{-3}$ cm$^{-3}$)  & 6.3, 1.1       & - \\
r$_c^{\star}$                                 & 1.73, 3.80     & - \\
$\beta^{\star}$                               & 1.04, 0.58     & - \\
$n_{e0}$ ($\cdot 10^{-3}$ cm$^{-3}$)          & 7.84           & 4.32 \\
r$_c$                                         & 0.925          & 1.98 \\
$\beta$                                       & 0.547          & 0.605 \\
$K_0$ (keV cm$^2$)                            & $136 \pm 25$   & $136 \pm 25$ \\
\hline                 
\end{tabular}
\end{center}
\begin{tablenotes}
\item \footnotesize{\textbf{Note:} asterisk in column 1 indicates parameters obtained by 
fitting the double $\beta$-model. The masses M$_{500}$ and M$_{gas}$ are the masses for an 
entire, spherically symmetric cluster with the profile of the named hemisphere. For details 
of the X-ray temperature see the profiles in Fig. 8 for A1307 and the temperature profile 
in OTA13 for A76. $K_0$ is the central entropy of the intracluster medium.}
\end{tablenotes}
%\end{threeparttable}
\end{table}

A plausible way to estimate the total cluster mass can be based on the gas-mass distribution. From the 
surface brightness profiles, we find gas masses inside the southern and northern parts of M$_{500}=1.5 
\cdot 10^{13}$ and M$_{500}=2.25 \cdot 10^{13}$ M$_\odot$, respectively, taking the mass only for each 
hemisphere. Adopting the assumption mentioned above, that the southern part of the cluster provides an 
impression of the undisturbed main cluster component, we find that the north hemisphere contributes with 
an additional gas mass of the order of $0.75 \cdot 10^{13}$ M$_\odot$, which is about
25\% of that of the main cluster component. This leads to a total mass estimate of M$_{500}=2.2 
\cdot 10^{14}$ M$_\odot$ plus 25\% resulting in M$_{500} \sim 2.75 \cdot 10^{14}$ M$_\odot$. We have 
taken the gas masses here to estimate the mass ratio of the two cluster components, since the gas 
mass can be obtained relatively straightforward from the surface brightness profiles and the 
geometry. The same exercise with the total mass would also include the assumption of hydrostatic 
equilibrium, which is certainly not clearly observed in this case. From this result, we may
conclude that we observe an ongoing merger with a mass ratio of about 1:8.

For the X-ray luminosity of A1307, in the [0.5-2] keV energy band we obtain a value of $L_{\rm{X},500} 
= 6.2$ and $7.4 \cdot 10^{43}$ erg s$^{-1}$ in the southern and northern regions, respectively. 
Based on the mass--luminosity relation proposed by \citet{Pra09}, we expect a total cluster mass 
of about M$_{500} = 2-3 \cdot 10^{14}$ M$_{\odot}$. On the other hand, using the M-T relation 
by \citet{Arn05}, for a X-ray temperature of 4.3 keV we estimate a total mass of about M$_{500} 
= 2.9 \cdot 10^{14}$ M$_{\odot}$ inside $r_{500}$. Although the mass obtained from the X-ray luminosity 
and temperature (to a lesser extent) of the ICM for cases of more relaxed clusters are in quite good
agreement, it is safer to use the gas mass and the very possible assumption that the ratio between 
the total mass and the gas mass is roughly constant. 

\section{Dynamical mass}
\label{sec:mass}

As has been argued previously, both A76 and A1307 are galaxy clusters showing some hints of dynamical 
unrelaxation. A76 shows a shallow, elongated and clumpy X-ray surface brightness, which coincides with a very 
spread spatial galaxy distribution and a high number of bright galaxies. On the other hand, A1307 
presents a more compact and bright X-ray emission, showing a clear asymmetry in the south-north direction. 
However, both galaxy clusters present a very Gaussian velocity distribution (see also Sect. 
\ref{sec:mass_recon}). In addition, several kinematical testing do not find evidences of clear 
substructures. 

Due to all these facts, we can assert that the interpretation of the dynamical state of A76 and 
A1307 depends a lot of the adopted point of view. That is, if one only considers the X-ray profiles 
(elongated shapes, moderate-high Tx and entropies) and forgets optical data, the clusters may be 
considered quite far from equilibrium. On the contrary, if one considers only optical data (mainly 
Gaussian velocity distributions, none important substructures detected, none BCG velocity offsets, no significant 
velocity gradients or anisotropies), the clusters are very close to the dynamical equilibrium. So, in 
order avoid a biased interpretation, we can only affirm that, despite the gas of the clusters can be disturbed, 
the galactic component is quite relaxed. In this scenario, it is possible to estimate virial dynamical 
masses and radii, and compute the dynamical masses of the whole clusters. 

Galaxies are tracers of the total gravitational potential of a cluster, so their velocities can be used
for estimating the dynamical mass of galaxy systems. Therefore, by utilising the velocity dispersion, 
$\sigma_\textrm{v}$, and its correlation with the virial mass, M$_{200}$, we estimate the dynamical 
mass of A76 and A1307. Scaling relations represent a frequently used method for estimating the dynamical 
masses of clusters based on their velocity dispersion. Important contributions in this sense have been
the relations proposed by \citet{Evr08}, \citet{Saro13} or \citet{Mun13}. However, we 
assume that found by \citet{Ferra20}, see Eq. (\ref{eq:ferra}). This last $\sigma_\textrm{v}-$M$_{200}$ 
relation is based and calibrated on \cite{Mun13} simulations, which considers dark matter particles 
and subhalos, galaxies, and AGN feedback, but goes further: \citet{Ferra20} relation also takes into 
account the physical and statistical effects in samples including small numbers of galaxy members. 

\begin{equation}
\label{eq:ferra}
    \frac{\sigma_\textrm{v}}{\textrm{km s}^{-1}} = \textrm{A} \Bigg[ \frac{h(z) \textrm{M}_{200}}{10^{15} 
    \textrm{M}_{\odot}} \Bigg] ^ \alpha \ \ \ ,
\end{equation}
where $\sigma_\textrm{v}$ is the observed velocity dispersion within a sphere of radius $r_{200}$,
and A and $\alpha$ are parameters fitted using \cite{Mun13} simulations, which take values of A$=1177$ km 
s$^{-1}$ and $\alpha=0.364$. 

Following this method, we find dynamical masses of M$_{200}=1.8 \pm 0.6 \cdot 10^{14}$ M$_{\odot}$ and 
$4.3 \pm 1.2 \cdot 10^{14}$ M$_{\odot}$ for A76 and A1307, respectively. To compare these values with 
those derived from X-ray emission we convert M$_{200}$ into M$_{500}$ following the relation given by 
\citet{Duf08}. That is, we rescale M$_{500}$ from M$_{200}$ assuming a concentration parameter $c_{200}=4$,
an appropriate value for clusters at $z<0.1$ and masses of M$_{200} \sim 10^{14}$ M$_{\odot}$, integrating 
a Navarro-Frenk-White (NFW) profile \citep{NFW97} and interpolating to obtain M$_{500}$. In this way, we
obtain, M$_{500}=1.1 \pm 0.3 \cdot 10^{14}$ M$_{\odot}$ and $2.7 \pm 0.7 \cdot 10^{14}$ M$_{\odot}$ for 
A76 and A1307, respectively. Table \ref{tab:properties} compiles these findings, some of them already 
advanced in in Sect. \ref{sec:velocity}. 

We can also determine the quasi-virialised regions by estimating the virial radius of each galaxy system.
The radius of this region can be obtained as that of a sphere with mass M$_{200}$ where the mass density
is 200 times the critical mass density of the Universe at the redshift of the system, 
$200\rho_c(\textrm{z})$. So, $M_{200}=100\,$r$_{200}^3 \textrm{H}(\textrm{z})^2/$G, and following this 
expression, we obtain r$_{200} = 1.2 \pm 0.2$ and $1.5 \pm 0.2$ $h_{70}^{-1}$ Mpc for A76 and A1307, 
respectively (see Table \ref{tab:properties}). It is important to note that, while we find an excellent 
agreement between the total dynamical mass and the mass obtained from X-ray emission for A1307, this 
comparison for A76 reveals a difference of about 70\%. We will discuss the origin of this discrepancy in 
Sect. \ref{sec:optical_xray_mass}.

\section{Mass profiles and velocity anisotropy with \textsc{MG-MAMPOSSt}} 

\label{sec:mass_recon}
In the following, we present a more detailed  reconstruction of the total mass profiles M$(r)$ and the 
velocity anisotropy profiles $\beta(r)$ of A1307 and A76 using data of velocities and projected positions 
of member galaxies. We perform a Bayesian analysis by means of  \textsc{MG-MAMPOSSt}  by \cite{Pizzuti23a}, 
a code for kinematic analyses of galaxy clusters based on the \textsc{MAMPOSSt} (Modelling Anisotropy and 
Mass Profile of Spherical Observed Systems) method of \cite{Mamon13}.
These algorithms simultaneously estimates the (total) gravitational potential and the orbit anisotropy 
profile of clusters assuming spherical symmetric and dynamical relaxation. They are based on the Jeans' 
equation:

\begin{equation}\label{eq:jeans}
\frac{\text{d} (\nu \sigma_r^2)}{\text{d} r}+2\beta(r)\frac{\nu\sigma^2_r}{r}=-\nu(r)\frac{\text{d} \Phi}{\text{d} r}\,\, ,
\end{equation}
where $\nu(r)$ is the number density profile of galaxies, $\sigma^2_r$ is the velocity dispersion along the 
radial direction and $\Phi$ is the total gravitational potential - related to the mass profile as $\text{d}\Phi(r)/
\text{d}r =G M(r)/r^2$, being $G$ the gravitational constant. The anisotropy is defined as
$\beta \equiv 1-(\sigma_{\theta}^2+\sigma^2_{\varphi})/2\sigma^2_r$, where $\sigma_{\theta}^2$ and $
\sigma^2_{\varphi}$ are the velocity dispersion components along the tangential and azimuthal directions, 
respectively. In spherical symmetry we have $\sigma_{\theta}^2=\sigma^2_{\varphi}$ and the expression of the 
anisotropy profile simplifies to $\beta = 1 -\sigma_{\theta}^2/\sigma_{r}^2$.

\textsc{MAMPOSSt} (and \textsc{MG-MAMPOSSt}) takes as input parametric expressions of the gravitational potential, 
the number density profile and the velocity anisotropy profile; by applying a Maximum Likelihood technique it 
determines the best set of parameters which describe the line-of-sight velocity distribution and projected 
positions of cluster member galaxies. The codes also require an assumption on the shape of the three-dimensional 
velocity distribution of the system. In our current implementation, a Gaussian model is used; however 
\cite{Mamon13} showed that \textsc{MAMPOSSt} works quite well also for non-Gaussian distributions. 
\textsc{MG-MAMPOSSt} extends the available models in \textsc{MAMPOSSt}, including new parametrisations of the 
gravitational potential in general modified gravity and possible dark energy scenarios (see \citealt{Piz21} 
for a description of the physics) as well as quite the generic ansatz for the anisotropy profiles.

% \subsection{Preparation of the dataset}
For both clusters, we consider the distribution of member galaxies in the projected phase-space 
$(R_{(i)},v_{z,(i)})$ - p.p.s. hereafter - where $R_{(i)}$ is the projected position of the $i$-th galaxy 
with respect to the centre of the cluster, assumed as the BCG position, also coincident with the X-ray 
peak of emission. $v_{z,(i)})$ indicates the velocity measured along the line of sight, in the rest-frame of 
the cluster. 

\begin{figure*}
\sidecaption
  \includegraphics[width=12cm]{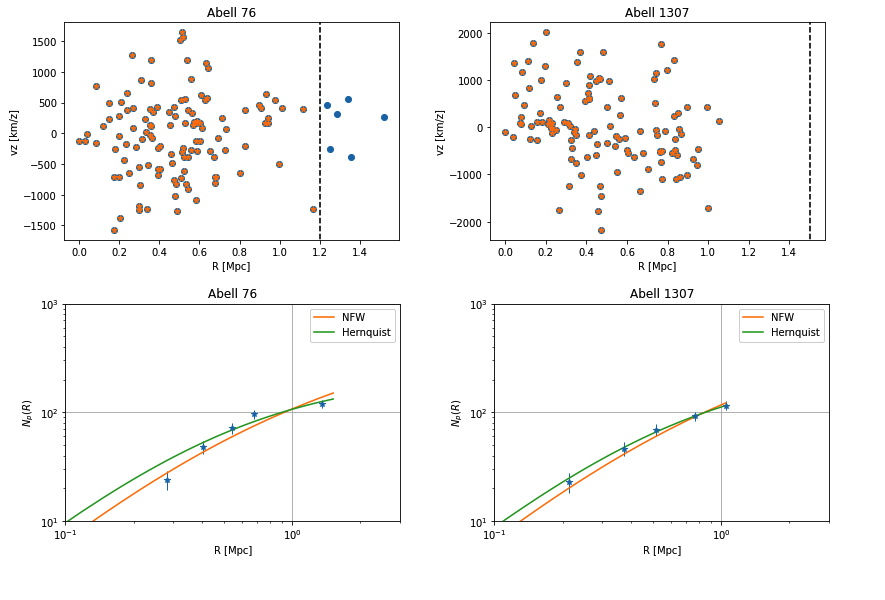}
     \caption{Projected phase spaces of Abell 76 (top left) and Abell 1307 (top right). The vertical dotted 
lines represent the estimated $r_{200}$ derived from the dynamical analyses. Bottom panels display the 
corresponding (projected) number density profiles (blue stars with error bars). The green and orange solid 
curves represent the best-fit profiles for a projected Hernquist and projected NFW models, respectively.}
     \label{fig:ps}
\end{figure*}

In the top panel of Figure \ref{fig:ps} we show the p.p.s. of A76, on the left, and  A1307 on the right. The 
vertical dashed line indicates the estimated value of $r_{200}$ discussed in the previous section. In the 
\text{MG-MAMPOSSt} fit we consider only members up to $R_\text{up}=r_{200}$ to ensure the validity of the 
Jeans equation; this will exclude a few galaxies in  A76 (blue points). We have nevertheless checked that 
changing the values of $R_\text{up}$ leads to negligible modification to the final results. In the case 
of  A1307, all the observed members are within $\sim 0.7\, r_{200}$.

We further exclude the innermost region $R<0.05 \,\text{Mpc}$, where the kinematics is dominated by the BCG. 
Our final sample consists of 112 galaxies for A1307 and 113 galaxies for A76.

Finally, we reconstruct the projected number density profiles of both cluster by fitting the projected space 
distribution $\{R_i\}$ with the a projected Navarro-Frenk-White (pNFW) and a projected Hernquist (pHer) model. We 
implement the Maximum Likelihood method of \cite{Sarazin1980}, which does not require the binning of data. 
Moreover, we assign at each member galaxy position a weight which is given by the inverse of the completeness at 
that radius (see Section~\ref{sec:optical_obs})

We find a slight, statistically irrelevant, preference for the pHer model over the pNFW for both cluster, with a $
\Delta$BIC $< 1$\footnote{The Bayesian Information Criterion (see e.g. \citealt{Mam19}) is defined as 
BIC$=-2\ln\mathcal{P}+N_\text{P}\ln\,N_\text{data}$, where $\mathcal{P}$ is the posterior distribution, 
$N_\text{P},\,N_\text{data}$ the number of free parameters and data points, respectively. A value of $\Delta$BIC$ 
\gtrsim 6$ means that the first model is highly disfavoured with respect to the second.} 
We obtain 
\begin{equation}
    r_\nu(\text{pNFW}) = 0.50^{+0.15}_{-0.12}\,\text{Mpc}\,\,\,\,\,\,\,\,\, r_\nu(\text{pHer}) = 1.08^{+0.25}_{-0.19}\,\text{Mpc}\,
\end{equation}
for A76 and 
\begin{equation}
    r_\nu(\text{pNFW}) = 0.56^{+0.40}_{-0.22}\,\text{Mpc}\,\,\,\,\,\,\,\,\, r_\nu(\text{pHer}) = 1.17^{+0.67}_{-0.37}\,\text{Mpc}\,
\end{equation}
for A1307. In both cases, the uncertainties are given at 95\% C.L. To assess the effect induced by a change in 
the value of the number density scale radius, we consider in the \textsc{MG-MAMPOSSt} $r_\nu$ as a free 
parameter assuming flat priors within the $2\sigma$ limits obtained by the Maximum Likelihood analysis.
In both cases, we consider four approaches for the total mass profile, namely the NFW model:
\begin{equation}
\textrm{M}_\text{NFW}(r) = \text{M}_{200} \frac{\log(1+r/r_\text{s})-r/r_\text{s}/(1+r/r_\text{s})}{\log(1+c_{200})-c_{200}/(1+c_{200})}\,,
\end{equation}
where, as before, $\text{M}_{200}=\left[ 100 H(z)^2/G\right] r_{200}^3$ is the mass at $r_{200}$, $r_\text{s}= r_{-2}$ 
is the scale radius at which the logarithmic slope of the density equals $-2$, and $c_{200}=r_{200}/r_{-2}$ is the concentration.

The Burkert profile \cite{Burkert01} is
\begin{equation}
    \textrm{M}_\text{Bur}(r) = \frac{\text{M}_{200}f(r/r_\text{s})}{f(r_{200}/r_\text{s})}\,,
\end{equation}
with $r_\text{s}\simeq 2/3 r_{-2}$, and 
\begin{equation}
    f(x) = \left\{ \ln\left[1+x^2\right]+2 \ln(1+x) -2\arctan (x)\right\}\,.
\end{equation}

On the other hand, the Hernquist model \cite{Hernquist01} is also defined as
\begin{equation}
    \textrm{M}_\text{Her}(r) = \text{M}_{200}\frac{(r_\text{s}+r_{200})^2}{r_{200}^2}\frac{r^2}{(r+r_\text{s})^2}\,,
\end{equation}
where $r_\text{s} = 2 r_{-2}$. While the Einasto model \cite{Einasto65} follows the expression:
\begin{equation}
    \textrm{M}_\text{Ein}(r) = \text{M}_{200}\frac{\mathcal{P}[3n,2n (r/r_\text{s})^{1/n}]}{\mathcal{P}[3n,2n (r_{200}/r_\text{s})^{1/n}]}\,,
\end{equation}
where $\mathcal{P}(\alpha,z) = \gamma(\alpha,z)/\Gamma(\alpha)$ is the regularised incomplete gamma function 
and $r_\text{s} = r_{-2}$. The exponent $n$ is assumed to be $n=5$, which has been shown to be a typical value 
for cluster-size halos in cosmological simulations (e.g. \citealt{Mam10}).

As for the velocity anisotropy, we consider a generalised Tiret (gT) model (see e.g. \citealt{Mam19,Biviano23}),
\begin{equation}
    \beta_\text{gT}(r)=\beta_0 + (\beta_\infty-\beta_0)\frac{r}{r+r_\beta}\,, 
\end{equation}
where $\beta_0$ and $\beta_\infty$ are the values of the central anisotropy and of the anisotropy at 
infinity, respectively. $r_\beta$ is a scale radius which is assumed to be equal to the radius $r_{-2}$ of 
the mass profile (Tied Anisotropy Number Density; \citealt{Mam10}). In \textsc{MG-MAMPOSSt}, we consider as 
parameters for the anisotropy profile the quantities $\mathcal{A}_X = (1-\beta_X)^{-1/2}$, which represent 
the ratios between radial and tangential velocity dispersion.

We explore the parameter space $(r_{200},r_\nu,r_\text{s},\mathcal{A}_0,\mathcal{A}_\infty)$ with a Markov-Chain 
Monte Carlo (MCMC) over $10^5$ points for both clusters for the models mentioned above. The best fit parameters 
with the 95\% confidence level uncertainties are displayed in Table \ref{tab:A76} for A76 and A1307. In all cases, 
a generalised Tiret (gT) profile has been adopted for the velocity anisotropy, and uncertainties are given 
at $2\sigma$ level.

\begin{table*}
\caption{\label{tab:A76} Results of different mass and number density models ansatz for 
the cluster A76 and A1307.}
\centering
\begin{tabular}{c|c|c|c|c|c|c}
\multicolumn{7}{c}{\textbf{Abell 76}} \\
\hline
\hline
\textbf{Model} &  $r_{200}\,\,[\text{Mpc}]    $	 & $r_{\nu}\,\, [\text{Mpc}]   $	 & $r_\text{s}\,\, [\text{Mpc}]     $	 & $\mathcal{A}_\infty  $	 & $\mathcal{A}_0       $	& $-\log P$ \\[0.2cm]
\hline
NFW+pHer & $1.48^{+0.35 }_{-0.29}$ & $0.99^{+0.19 }_{-0.11}$ & $0.95^{+2.01 }_{-0.95}$ & $1.48^{+1.88 }_{-1.02}$ & $1.92^{+2.44 }_{-1.47}$ &  892.80\\[0.2cm]
Bur+pHer & $1.43^{+0.51 }_{-0.69}$ & $1.00^{+0.21 }_{-0.12}$ & $0.73^{+1.87 }_{-0.73}$ & $1.67^{+1.87 }_{-1.19}$ & $2.01^{+2.37 }_{-1.56}$ & 892.81 \\[0.2cm]
Her+pHer & $1.47^{+0.33 }_{-0.28}$ & $1.00^{+0.19 }_{-0.11}$ & $1.37^{+1.95 }_{-1.33}$ & $1.48^{+1.93 }_{-1.02}$ & $1.90^{+2.45 }_{-1.44}$& 892.85 \\[0.2cm]
Ein5+pHer & $1.48^{+0.34 }_{-0.29}$ & $1.00^{+0.19 }_{-0.11}$ & $1.04^{+2.03 }_{-1.04}$ & $1.50^{+1.87 }_{-1.04}$ & $1.90^{+2.42 }_{-1.45}$& 892.89 \\[0.2cm]
NFW+pNFW & $1.40^{+0.42 }_{-0.32}$ & $0.45^{+0.12 }_{-0.07}$ & $0.89^{+1.97 }_{-0.89}$ & $1.60^{+1.94 }_{-1.13}$ & $1.97^{+2.38 }_{-1.51}$& 894.65 \\[0.2cm]
Bur+pNFW & $1.37^{+0.42 }_{-0.60}$ & $0.45^{+0.12 }_{-0.07}$ & $0.53^{+1.20 }_{-0.51}$ & $1.68^{+1.88 }_{-1.20}$ & $1.98^{+2.37 }_{-1.53}$& 894.62 \\[0.2cm]
Her+pNFW & $1.41^{+0.34 }_{-0.29}$ & $0.45^{+0.12 }_{-0.07}$ & $1.32^{+1.93 }_{-1.27}$ & $1.61^{+1.88 }_{-1.14}$ & $1.87^{+2.49 }_{-1.42}$& 894.69 \\[0.2cm]
Ein5+pNFW & $1.40^{+0.38 }_{-0.33}$ & $0.45^{+0.12 }_{-0.07}$ & $0.96^{+1.97 }_{-0.96}$ & $1.62^{+1.90 }_{-1.16}$ & $1.89^{+2.42 }_{-1.44}$& 894.70 \\
\hline 
\multicolumn{7}{c}{\textbf{Abell 1307}} \\
\hline
\hline
\textbf{Model} &  $r_{200}\,\,[\text{Mpc}]    $	 & $r_{\nu}\,\, [\text{Mpc}]   $	 & $r_\text{s}\,\, [\text{Mpc}]     $	 & $\mathcal{A}_\infty  $	 & $\mathcal{A}_0       $	& $-\log P$ \\[0.2cm]
\hline
NFW+pHer & $1.91^{+0.50 }_{-0.43}$ & $1.00^{+0.36 }_{-0.21}$ & $1.00^{+1.54 }_{-0.98}$ & $1.47^{+2.54 }_{-1.01}$ & $1.79^{+3.01 }_{-1.36}$& 908.78 \\[0.2cm]

Bur+pHer & $2.06^{+1.14 }_{-0.69}$ & $1.00^{+0.35 }_{-0.21}$ & $1.05^{+1.55 }_{-1.03}$ & $1.68^{+2.48 }_{-1.23}$ & $2.35^{+2.94 }_{-1.87}$& 908.80 \\[0.2cm]

Her+pHer & $1.87^{+0.45 }_{-0.41}$ & $1.00^{+0.37 }_{-0.21}$ & $1.34^{+1.40 }_{-1.16}$ & $1.35^{+2.51 }_{-0.89}$ & $1.85^{+3.21 }_{-1.42}$& 908.81 \\[0.2cm]

Ein5+pHer & $1.90^{+0.48 }_{-0.42}$ & $1.03^{+0.22 }_{-0.15}$ & $1.05^{+1.52 }_{-1.02}$ & $1.47^{+2.65 }_{-1.01}$ & $1.74^{+3.15 }_{-1.30}$& 909.00 \\[0.2cm]

NFW+pNFW &  $1.83^{+0.53 }_{-0.44}$ & $0.44^{+0.21 }_{-0.12}$ & $1.04^{+1.53 }_{-1.01}$ & $1.54^{+2.50 }_{-1.08}$ & $1.98^{+3.05 }_{-1.55}$& 909.19 \\[0.2cm]

Bur+pNFW & $1.72^{+1.02 }_{-1.30}$ & $0.44^{+0.23 }_{-0.13}$ & $1.11^{+1.55 }_{-1.07}$ & $1.79^{+2.42 }_{-1.34}$ & $2.56^{+2.86 }_{-2.04}$& 909.05 \\[0.2cm]

Her+pNFW & $1.81^{+0.45 }_{-0.39}$ & $0.43^{+0.21 }_{-0.12}$ & $1.37^{+1.39 }_{-1.14}$ & $1.54^{+2.56 }_{-1.08}$ & $1.85^{+3.04 }_{-1.42}$& 909.19 \\[0.2cm]

Ein5+pNFW & $1.83^{+0.50 }_{-0.43}$ & $0.43^{+0.18 }_{-0.11}$ & $1.06^{+1.51 }_{-1.01}$ & $1.57^{+2.63 }_{-1.12}$ & $1.89^{+3.08 }_{-1.45}$& 909.18 \\
\hline
\end{tabular}
\end{table*}

\begin{figure}
\centering
\includegraphics[width=\columnwidth]{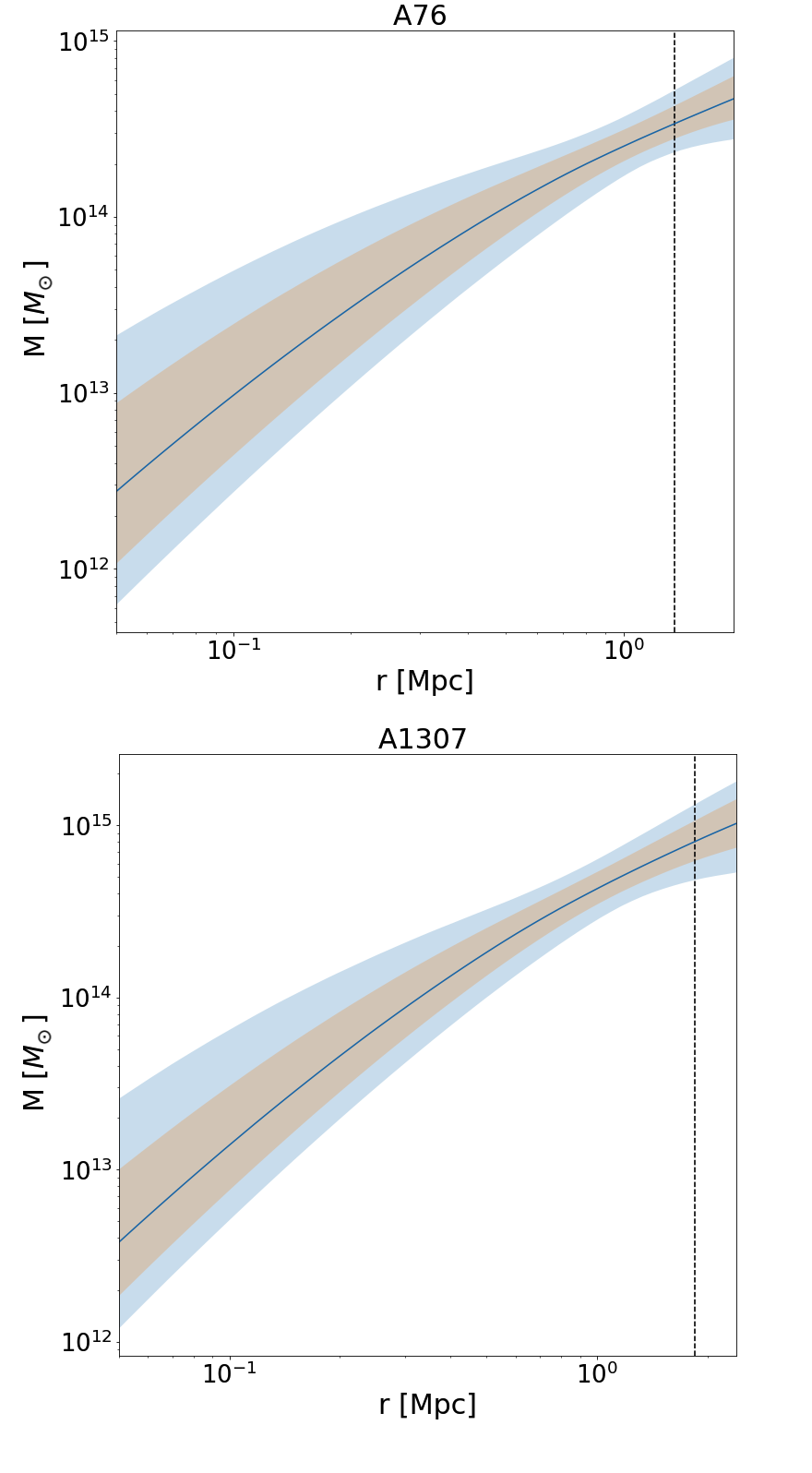}
\caption{\label{fig:massp} Best-fit NFW mass profile for A76 (top) and A1307 (bottom). The inner and outer 
shaded area represent the 68\% and 95\% confidence regions, respectively. The black dotted vertical lines 
indicate the best-fit value of $r_{200}$.}
\end{figure}

We find that all the best fit mass profiles are statistically comparable, with no strong evidence for a 
preferred ansatz. We thus choose as a reference the NFW model with a pHer for the number density profile 
(NFW+pHer). The NFW mass distribution as a function of the radius is shown in Figure \ref{fig:massp} for 
A76 (top) and A1307 (bottom).

\begin{figure}
\centering
\includegraphics[width=1.0\linewidth]{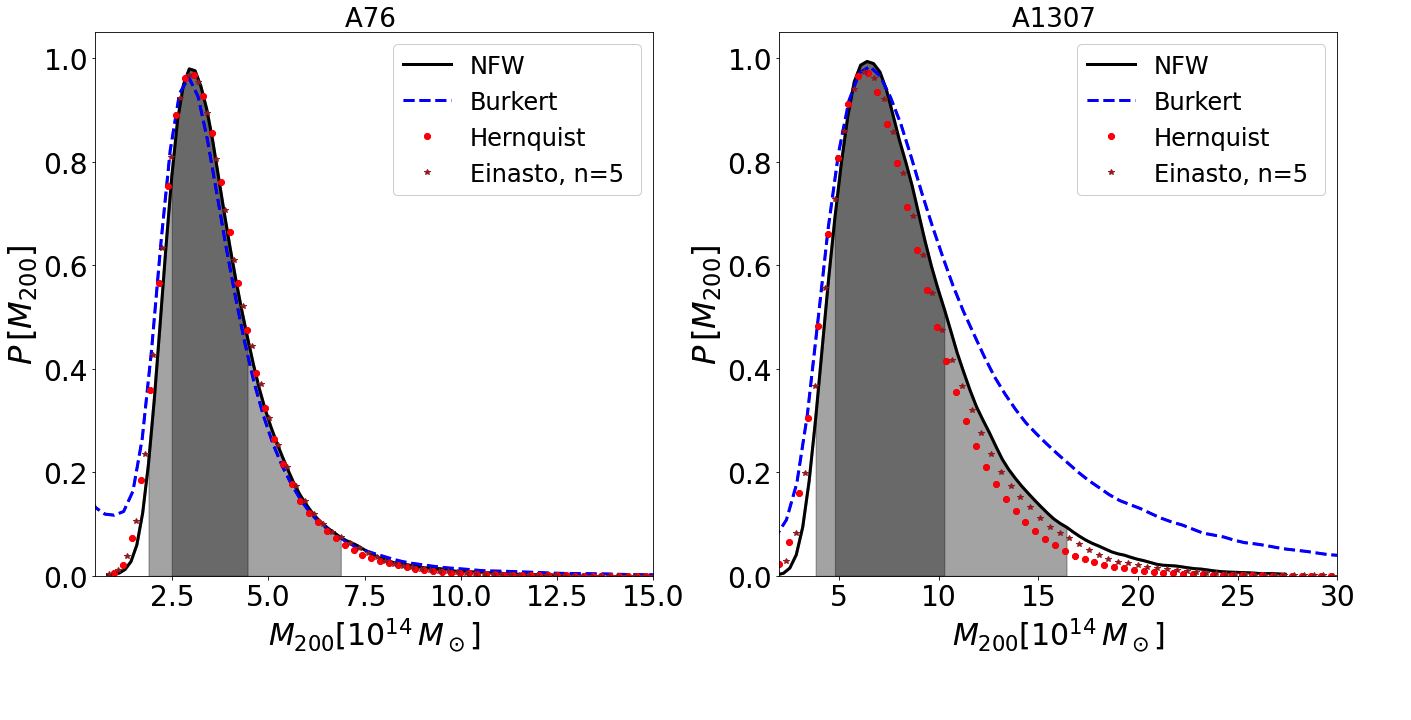}
\caption{\label{fig:M200} Marginalised posterior of M$_{200}$ obtained by fitting the phase space with a 
NFW model for the mass profile (solid black curve), a Burkert profile (dashed blue line), Hernquist (filled red 
circles), and Einasto with $n=5$ (brown stars). The shaded areas refer to the 68\% (dark grey) and 95\% 
(light grey) confidence intervals.}
\end{figure}

We quote, at 68\% confidence level,
\begin{equation}
    r_{200}=1.48^{+0.35}_{-0.29}\,\text{Mpc} \,\,\,  \text{(A76)}\,\,\,\,\, ; \,\,\, r_{200}=1.91^{+0.50}_{-0.43}  \,\,\, \text{(A1307)}\,,
\end{equation}
which correspond to a mass
\begin{equation}
    \text{M}_{200}=2.9^{+1.6}_{-0.6}\,\text{M}_\odot \,\, \text{(A76)}\,\,\,\, ; \,\,\,\,\, \text{M}_{200}=7.7^{+2.8}_{-3.0}\times 10^{14}\,\text{M}_\odot  \,\, \text{(A1307)}\,.
\end{equation}
Figure \ref{fig:M200} further displays the posterior marginalised distribution of $M_{200}$ for the different 
mass models analysed assuming a pHer profile for the number density. The curves are overall in good agreement, 
with the Burkert model showing the largest uncertainties. 

We note that values of M$_{200}$ are larger than what we find from the dynamical mass-velocity dispersion relation, 
Eq. (\eqref{eq:ferra}); in order to compare with the X-ray luminosity, we can compute M$_{500}$ directly from the 
NFW best fit profile. We obtain, at 1$\sigma$, M$_{500} = 2.1^{+0.5}_{-0.4}\times 10^{14}$ for A76 and 
M$_{500} = 4.5^{+1.1}_{-1.0}\times 10^{14}$ for A1307. Surprisingly, the results of the \textsc{MG-MAMPOSSt} fit 
show a better agreement with the X-ray estimate of A76 rather than for A1307, even if the value are still in 
agreement within 1$\sigma$. The masses inferred with \textsc{MG-MAMPOSSt} are further reported in 
Table~\ref{tab:properties}.

\begin{figure}
\includegraphics[width=0.9\columnwidth]{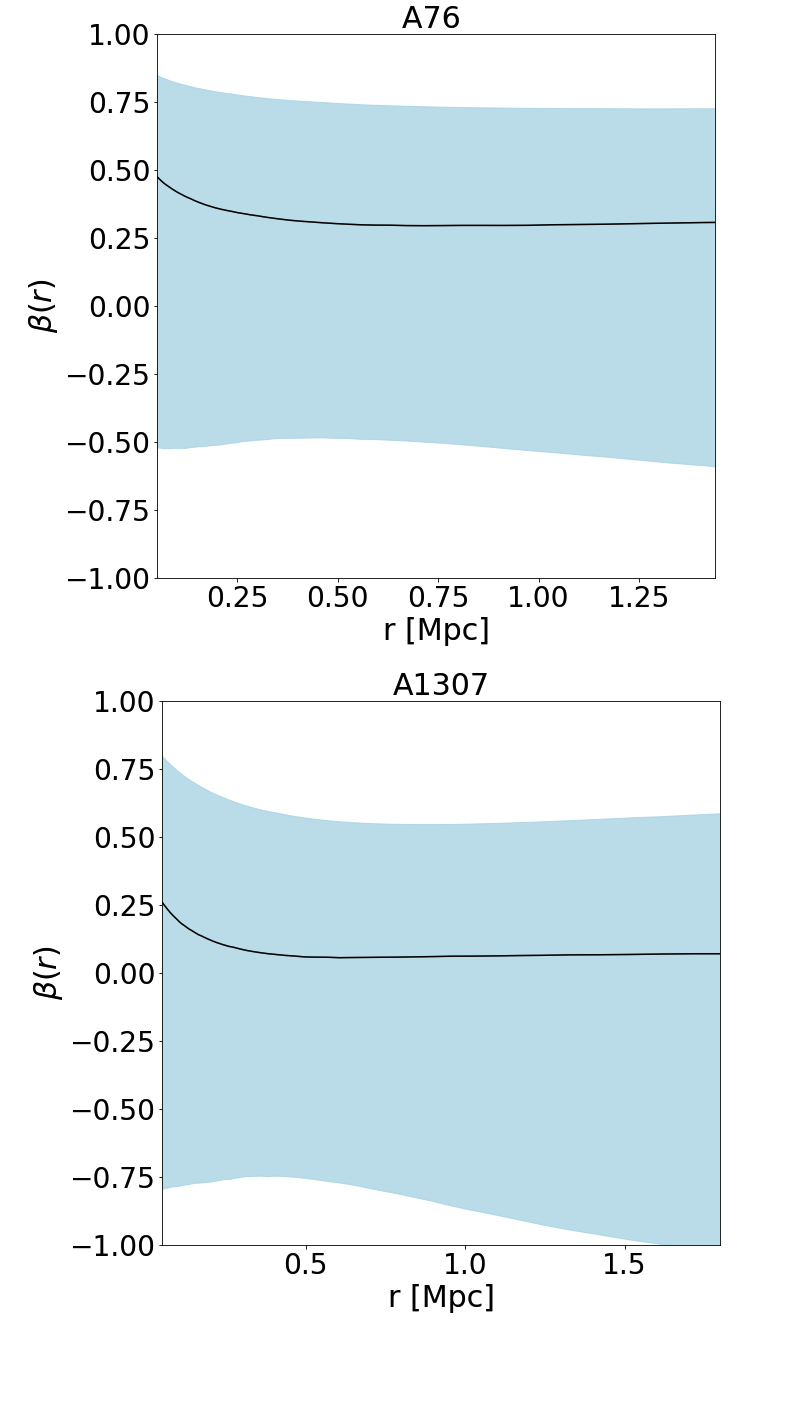}
\caption{\label{fig:anis} Radial profiles for the best-fit velocity anisotropy of A76 (top) and A1307 (bottom), parametrised as a gT model. The shaded areas refer to the 68\% confidence intervals.}
\end{figure}

The radial profiles for the velocity anisotropy are shown in Figure \ref{fig:anis}; given the large uncertainties one 
can only state a qualitative behaviour of the orbits for the two clusters. In particular, A76 seems to slightly 
favour more radial orbits towards the whole cluster range, while the best fit of A1307 is in agreement with fully 
isotropic orbits down to the very cluster centre.

\section{Explaining optical and X-ray properties}
\label{sec:optical_xray_mass}

From the analyses of the X-ray data (see Sect. \ref{sec:x-ray}), A76 shows a very elongated configuration, 
very shallow and extended in the east-west direction, with moderate individual peaks on each of its five 
brightest galaxies, making its X-ray quite clumpy. The elongated X-ray emission and 2D galaxy distribution 
in A76 suggest that its dynamical state is not completely relaxed. However, after applying several tests on 
the velocities and positions of the optical and X-ray emission, we have not been able to identify any clear 
subcluster as responsible of a merge event. Beside the BCG, we identify up to four bright galaxies more that 
could be associated to small galaxy groups. One or several recent minor mergers of these groups could be the 
responsible mechanism of the so clumpy and elongated X-ray emission in the E-W direction. We find that 
M$_{500}$ derived from X-ray temperature, is about $1.9 \cdot 10^{14}$ M$_{\odot}$, in agreement with the 
\textsc{MG-MAMPOSSt} analysis but slightly higher value than we found for the dynamical mass, M$_{500,dyn} 
\sim 1.1\cdot 10^{14}$ M$_{\odot}$. A plausible explanation is that the T$_X$ is somehow enhanced, maybe due 
to the geometry of the three-dimensional gas distribution or, as it is also pointed out by OTA13, due to recent 
minor mergers, that could disturb the dynamical state of the cluster. In a similar way, the clumpy distribution 
of galaxies can bias the mass estimation based on \textsc{MG-MAMPOSSt}.

A1307 presents a more regular scenario where X-ray emission is more compact. However, the X-ray surface 
brightness presents a clear elongation following the S-N direction, almost perpendicular to the velocity gradient.
On the other hand, M$_{500,\rm{X}}$, derived from X-ray luminosity and temperature, fits quite well with the 
dynamical mass. Only, the elongation of the X-ray emission and 2D galaxy density distribution suggests that 
the dynamical state of A1307 may be associated to minor merger event in the recent past of the cluster. Moreover, 
the mass profile obtained with the \textsc{MG-MAMPOSSt} fit seems to prefer larger values of M$_{200}$ with 
respect to the dynamical and X-ray estimations. However, it is important to remark that this fact could be 
related to the unknown three-dimensional shape and the direction of the ellipsoidal axes of the cluster. 
Indeed, as shown in, for example, \cite{Mamon13} and \cite{Pizzuti20}, \textsc{MAMPOSSt} tends to overestimate 
the total mass when the line of sight is aligned with the major axis of a cluster; in addition, we note that 
the p.p.s of A1307 covers only up to $R\sim0.7 \,r_{200}$, which can produce biases in the 
\textsc{MG-MAMPOSSt} fit.

\section{Summary and conclusions}
\label{sec:conclusions}

We perform a detailed study on the dynamical state of two galaxy clusters, Abell 76 and 1307, with different 
X-ray phenomenology. Our study relies on recent spectroscopic data obtained with the 3.5m TNG telescope, along 
with complementary NED and SDSS-DR16 spectroscopic redshifts within a region of $\sim 1$ and $0.7r_{200}$ for
A76 and A1307, respectively. We selected 122 and 116 cluster members to analyse the velocity field in each cluster.
These studies do not reveal any significant deviations from Gaussianity neither in A76 nor in A1307. Our results 
on the total mass of the cluster, based on the $\sigma-$M relation, the X-ray emission, the \textsc{MG-MAMPOSSt} 
analysis, and even the mass obtained from SZ-{\it Planck} signal, we obtain a mean M$_{500}$ of $1.7 \pm 0.6 \cdot 
10^{14}$ M$_{\odot}$ and $3.5 \pm 1.3 \cdot 10^{14}$ M$_{\odot}$ for A76 and A1307, respectively. 

For A76, we find that the X-ray surface brightness morphology and the projected galaxy distributions indicate 
that the clusters dynamics is not completely relaxed. Thus, we basically confirm the findings by OTA13. They 
suggest that A76 is at an early stage of cluster formation and that gas compression caused by potential 
confinement is lagging behind the gas heating. Our spectroscopic and photometric analyses also indicate that 
A76 is a young cluster in process of formation, that may be in the first stages 
of relaxation, further encouraged by a slight evidence radial orbits. 

Similarly, a large population of early-type galaxies, and the agreement between X-ray and dynamical magnitudes 
in A1307 suggest that this cluster may be more relaxed than A76, with an indication of more isotropic orbits as found 
by the \textsc{MG-MAMPOSSt} fit. However, its elongated X-ray emission reveals that the gas of this cluster has 
been disturbed by a recent minor merger. In addition, the mass profile obtained from the fit of the p.p.s. prefers 
larger M$_{200}$ with respect to what found from X-ray and $\sigma-$M relation, a possible indication 
of an elongated feature along the line of sight.

To summarise, A76 and A1307 represent two clusters with different masses and X-ray emissions but showing similar 
morphologies with a disturbed appearance in the X-ray images but no pronounced signatures of substructure in the 
galaxy velocity distribution in the line of sight. A76 is a young and not consolidated cluster, with only a few 
and very poor galaxy clumps that seem to be assembling, with a low mass and low X-ray surface brightness. Several 
features support this interpretation of a young dynamical system. The X-ray emission is very dispersed and it is 
difficult to explain this structure, for example, by a post-merger. We see several bright galaxies of which at 
least four have compact X-ray halos without any signs of distortions like stripping tails. Also the indication 
of a preference of radial orbits and a large fraction star forming emission line galaxies are in line with this 
interpretation. An remaining enigma is the high central entropy. If we compare this entropy value to other 
clusters for example in the REXCESS sample \citep{Pra09}, we find that this value corresponds to typical values 
in non-cool core clusters at a radius of 0.1 – 0.2 $r_{500}$. Since A76 does not have a real core, this value may 
not be so surprising. The typical low entropy value in the core may only be formed, when a dense, compact core 
is established. To further clarify the nature of these kind of systems, with so unusually low X-ray emission, 
many more dynamical studies have to be developed on similar clusters. On the other hand, A1307 represents the 
case of a more regular cluster, more massive and relaxed in comparison with A76. However, we cannot completely 
discard the possibility that A1307 has recently suffered a minor merger in the plane of the sky, involving a 
very low-mass and very poor galaxy clump, that remains undetectable in the velocity--space plane, but could 
slightly have disturbed in some way the X-ray emission.

With this work we want to enhance the importance of carrying out dynamic studies in galaxy clusters from a 
multi-wavelength approach. This is the natural way to find appropriate scenarios to disentangle the dynamical 
state the different components (galaxies and gas) of galaxy systems.

\begin{acknowledgements}

This project has been mainly funded by the Spanish Ministerio de 
Ciencia, Innovaci\'on y Universidades through grant PID2021-122665OB-I00.
R. Barrena acknowledges support by the Severo Ochoa 2020 research programme of 
the Instituto de Astrof\'{\i}sica de Canarias. G. Chon acknowledges support by 
the DLR under the grant n$^\circ$ 50 OR 2204. H. B\"ohringer acknowledges support 
from the Deutsche Forschungsgemeinschaft through the Excellence cluster 
"Origins".\\

This article is based on observations made with the Italian {\it Telescopio Nazionale 
Galileo} operated by the Fundaci\'on Galileo Galilei of the INAF (Istituto Nazionale 
di Astrofisica). This facility is located at the Spanish del Roque de los 
Muchachos Observatory of the Instituto de Astrof\'{\i}sica de Canarias on the 
island of La Palma. \\

Funding for the Sloan Digital Sky Survey (SDSS) has been provided by the Alfred
P. Sloan Foundation, the Participating Institutions, the National Aeronautics
and Space Administration, the National Science Foundation, the U.S. Department
of Energy, the Japanese Monbukagakusho, and the Max Planck Society. 

\end{acknowledgements}

\bibliographystyle{aa}

\end{document}